%% file: ms.tex
\documentclass[dvips]{aa} % for a referee version

\usepackage{amsmath}
\usepackage{amssymb}
\usepackage{graphicx}
\usepackage{txfonts}
\usepackage{natbib}
\def\nemb{43}
\def\nsources{61}
\def\nedge{10}
\def\nunknown{8}

\newcommand{\cmrk}{{\bf$\checkmark$}}
\newcommand{\micron}{$\mu$m}

\newcommand{\osia}{{\sc osia}}
\newcommand{\RA}[4]{$\rm#1^h\,#2^m\,#3^s#4$}
\newcommand{\Dec}[4]{$#1\degr\,#2\arcmin\,#3\arcsec#4$}

\newcommand{\spitzer}{\textit{Spitzer}}
\newcommand{\herschel}{\textit{Herschel}}

\newcommand{\neii}{[Ne\,{\sc ii}]}
\newcommand{\neiii}{[Ne\,{\sc iii}]}
\newcommand{\niii}{[Ni\,{\sc ii}]}
\newcommand{\feii}{[Fe\,{\sc ii}]}
\newcommand{\silii}{[Si\,{\sc ii}]}
\newcommand{\si}{[S\,{\sc i}]}
\newcommand{\oi}{[O\,{\sc i}]}
\newcommand{\cii}{[C\,{\sc ii}]}
\newcommand{\siii}{[S\,{\sc iii}]}
\newcommand{\fei}{[Fe\,{\sc i}]}
\newcommand{\cli}{[Cl\,{\sc i}]}
\newcommand{\htwo}{H$_2$}
\newcommand{\water}{H$_2$O}
\newcommand{\acetylene}{C$_2$H$_2$}
\newcommand{\cotwo}{CO$_2$}
\newcommand{\tex}{$T_\mathrm{ex}$}
\newcommand{\cshocks}{\hbox{$C$-shocks}}
\newcommand{\jshocks}{\hbox{$J$-shocks}}
\newcommand{\cshock}{\hbox{$C$-shock}}
\newcommand{\jshock}{\hbox{$J$-shock}}

\newcommand\T{\rule{0pt}{2.6ex}}

\newskip\tnotemarkskip
\newcommand\tablenotemark[1]{%
\rlap{$^{\mathrm #1}$}\hskip\tnotemarkskip\ignorespaces% Fixed: the space after notemark
}%

\def\@tablenotetext#1#2{%
 \vspace{.5ex}%
 {\noindent\llap{$^{#1}$}#2}%
}%
\newcommand\tablenotetext[2]{\strut\newline%
\parbox{0.025\columnwidth}{\hfill}%
\@tablenotetext{#1}{#2}%
}%

\begin{document}

\authorrunning{Fred Lahuis et al.}
%\titlerunning{c2d IRS Spectra of Gas-phase Emission toward obscured low mass young stars}
\titlerunning{c2d IRS spectra of embedded sources: gas-phase emission lines}

\title{c2d \textit{Spitzer} IRS spectra of embedded low-mass young stars: gas-phase emission lines}
\author{Fred~Lahuis\inst{1,2},
        Ewine~F.~van~Dishoeck\inst{2,3},
        Jes~K.~J{\o}rgensen\inst{4},\\
        Geoffrey~A.~Blake\inst{5},
        Neal~J.~Evans~II\inst{6}%,
	%Klaus~M.~Pontoppidan\inst{5} 
       }
\offprints{Fred Lahuis\\ \email{F.Lahuis@sron.nl}}

\institute{SRON Netherlands Institute for Space Research,
           P.O. Box 800, 9700 AV Groningen, The Netherlands
     \and
           Leiden Observatory, Leiden University, P.O. Box 9513,
           2300 RA Leiden, The Netherlands
     \and
           Max-Planck-Institut f\"ur extraterrestrische Physik (MPE),
           Giessenbachstraat 1, 85748 Garching, Germany
     \and
          Centre for Star and Planet Formation,
          Natural History Museum of Denmark,
          University of Copenhagen,\\
          {\O}ster Voldgade 5-7, DK-1350 Copenhagen K, Denmark
%          Argelander Institut f\"ur Astronomie, abt. Radioastronomie,
%          Auf dem H\"ugel 71, 53121 Bonn, Germany
     \and
          Division of Geological and Planetary Sciences 150-21, 
          California Institute of Technology, Pasadena, CA 91125
     \and
          The University of Texas at Austin, Dept. of Astronomy, 
          1 University Station C1400, Austin, Texas 78712--0259
}

\abstract{%
%context
A survey of mid-infrared gas-phase emission lines of \htwo, \water{}
and various atoms toward a sample of \nemb{} embedded low-mass young
stars in nearby star-forming regions is presented. The sources are
selected from the \spitzer{} "Cores to Disks" (c2d) legacy
program. 
%EvD: leave out; not needed in abstract
%The sample selection is driven by the presence of silicate
%and/or ice absorption features in the IRS spectra, and thus consists
%of both embedded young stars and edge-on and/or extincted disks.  
}%
%aims
{% 
The environment of embedded protostars is complex both in its physical structure (envelopes, outflows, jets, protostellar disks) and the physical processes (accretion, irradiation by UV and/or X-rays, excitation through slow and fast shocks) which take place. The mid-IR spectral range hosts a suite of diagnostic lines which can distinguish them. A key point is to spatially resolve the emission in the \spitzer-IRS spectra to separate extended PDR and shock emission from compact source emission associated with the circumstellar disk and jets.}%
%methods
{% 
An optimal extraction method is used to separate both spatially unresolved (compact, up to a few hundred AU) and spatially resolved (extended, thousand AU or more) emission from the IRS spectra. The results are compared with the c2d disk sample and literature PDR and shock models to address the physical nature of the sources.}%
%results
{%
Both compact and extended emission features are observed. Warm (\tex\,few hundred\,K) \htwo, observed through the pure rotational \htwo{} S(0), S(1) and S(2) lines, and \si{} 25 $\mu$m emission is observed primarily in the extended component. \si{} is observed uniquely toward truly embedded sources and not toward disks. On the other hand hot (\tex\,$\gtrsim700$\,K) \htwo, observed primarily through the S(4) line, and \neii{} emission is seen mostly in the spatially unresolved component. \feii{} and \silii{} lines are observed in both spatial components. Hot \water{} emission is found in the spatially unresolved component of some sources.}%
%conclusions
{%
The observed emission on $\geq$1000 AU scales is characteristic of PDR emission and likely originates in the outflow cavities in the remnant envelope created by the stellar wind and jets from the embedded young stars. Weak shocks along the outflow wall can also contribute. The compact emission is likely of mixed origin, comprised of optically thick circumstellar disk and/or jet/outflow emission from the protostellar object.}

\maketitle

%\clearpage
%\newpage

\section{Introduction}
\label{sec:introduction}

Understanding the nature of deeply embedded protostars is a major challenge in the study of low-mass star formation. In its earliest stages, a young star is embedded in its natal envelope of cold gas and dust, prohibiting direct detection with optical and near-IR instruments. During the evolution of such protostars the envelope is dissipated partly due to the ongoing accretion onto the central star and disk and partly due to the action of the outflows driven by such stars. This embedded phase lasts only a short time, $\sim6\times10^5$ yr \citep{evans09}, but is critical for the subsequent evolution of the star: it is the phase in which the final mass of the star, the overall mass of the circumstellar disk (and thus its ability to form planets), and the initial physical and chemical structure of the disk are determined.

The envelope surrounding young stars is subjected to highly energetic phenomena, in particular X-rays and UV radiation and streams of gas ejected from the star-disk boundary region. Both the slow and high velocity gas can shape the protostellar envelopes on a variety of spatial scales. The related physical processes, outlined below, have a strong impact on the physical structure and the molecular and atomic contents of the surrounding material. Mid-infrared spectroscopy, the central focus of this paper, provides a powerful tool to characterize and quantify the nature and energetics of these interactions as well as the composition and temperature structure of the material involved.

Based on previous infrared and submillimeter data, a general picture has emerged of the structure of embedded young stellar objects (YSOs) \citep[e.g.,][]{young04,joergensen02,joergensen05,chiang08,tobin10}. Take as an example the isolated source HH\,46\,IRS (IRAS08242-5050) imaged with the Spitzer Space Telescope \citep[see][and references cited]{noriega04,velusamy07} and studied in detail with submillimeter molecular lines as well as dust continuum emission \citep[see summary in][]{vankempen09b}. The young star is surrounded by a roughly spherically symmetric envelope with a mass of $\sim$5\,M$_\odot$ and a power-law density structure, with $n\mathrm{(H_2)}\,\approx\,10^6\,\mathrm{cm}^{-3}$ at 1000\,AU from the star, decreasing to $\sim 10^5$\,cm$^{-3}$ at $\sim$7000\,AU. The infrared images reveal prominent red- and blue-shifted outflow lobes traversing the envelope with an opening angle up to 110$^o$ and continuing out to 0.2\,pc. These outflow lobes are produced by jets and/or wide angle winds propagating at speeds up to 200\,km\,s$^{-1}$, which interact with the surroundings and create a shocked shell of material that is entrained and propelled outward at some speed \citep[e.g.,][]{bally07,ray07,shang07}. The most prominent example of this interaction is the bow shock at the tip of the outflow. In addition, blobs of shocked warm gas and dust are seen at various locations along the outflow axis, which may be internal shocks in the (variable) wind or jet itself. Similar physical structures are expected for other embedded YSOs, although the densities at 1000\,AU are up to an order of magnitude lower for sources with lower mass envelopes \citep{joergensen02}.

The outflow cavities create a low-extinction pathway for ultraviolet (UV) photons produced in the accreting gas column at the star-disk boundary layer to escape out to much further distances \citep{spaans95,bruderer09a,bruderer09b} than possible in a purely spherical symmetry \citep[e.g.][]{stauber04}. For a UV luminosity of 0.1\,L$_\odot$ typical of low-mass YSOs, the enhancement of the UV radiation field with respect to the general radiation field is about \hbox{$G_o$\,$\approx$\,$10^4$} at 100\,AU. Moreover, UV radiation is created along the outflow by the shocks themselves, especially by the higher velocity \hbox{$J$-type} shocks \citep{hollenbach89,neufeld89a}. Evidence for UV enhancements $G_o$ of a few hundred out to 0.1\,pc have recently been inferred from observations of narrow high-$J$ CO submillimeter emission lines surrounding the outflow cavity walls \citep[][see their Fig.\,13]{vankempen09b,vankempen09c}. Thus, the Spitzer IRS slit of about $5-10''$ will pick up emission from gas with a range of densities from $10^4-10^7$\,cm$^{-3}$, exposed to UV radiation with \hbox{$G_o\approx 10^2-10^4$}, and subjected to shocks with velocities up to several hundred km\,s$^{-1}$ for typical cloud distances of \hbox{$100-400$\,pc}.

%*** Move your Fig. 11 upward to start of manuscript? ***

The consequences of UV radiation impacting on dense gas have been studied for decades through models of so-called Photon-Dominated Regions (or PhotoDissociation Regions, PDR) \citep{tielens85a,tielens85b,hollenbach91,hollenbach99}. In the PDR layers closest to the UV source most molecular material is dissociated and the atomic species with ionization potential below 13.6\,eV are ionized. This layer is characterized by bright fine structure lines such as \oi, \fei, \feii{} and \silii. \si{} is never strong as most of the sulfur is in the form of S$^+$ in the warm PDR. At deeper layers, H$_2$ can survive, yet remains highly excited, resulting in strong pure rotational \htwo{} lines \citep[e.g.][applied to regions of massive star-formation]{kaufman06}.

Shocks impacting the surrounding envelope can be divided into two types: $J-$ and $C-$type shocks. The so-called dissociative \jshocks{} \citep{draine93,hollenbach97b,walmsley05} heat the gas up to $10^5$ K. Infrared emission comes from the dense hot post-shock gas in which most of the molecular material is still dissociated and a large fraction of the atoms are ionized resulting in strong emission lines of singly and doubly ionized atoms (e.g., \feii{} 26 and 35 $\mu$m, \niii{} 6.6 $\mu$m). This includes atoms which cannot be directly photoionized by photons with energies $<$13.6 eV such as the \neii{} 12.8 $\mu$m line \citep{hollenbach89,hollenbach09}. Further downstream, molecular hydrogen forms in the cooling gas and strong high-$J$ pure rotational line emission is expected. \neii{} emission is strong only in high velocity shocks \hbox{($v_\mathrm{s}>70$\,km\,s$^{-1}$)}, while the strength of other fine structure lines (e.g. \si{} 25 $\mu$m, \fei{} 24 $\mu$m and \cli{} 11.4 $\mu$m) does not strongly depend on the shock velocity \citep[see][for details]{hollenbach89}. Note that these fast shocks imply travel times through the envelope of only ($<10^4$\,yrs), unless the material is significantly slowed down by the interaction. For the so-called non-dissociative \cshocks{} \citep[e.g.][]{neufeld93,kaufman95,kaufman96}, the temperatures never become high enough to dissociate molecules; these shocks show strong lines from excited \htwo, CO, \water{} and OH \citep[e.g.][]{kaufman96}.

To date, PDR and shock models have mostly been applied to the general interstellar medium (ISM) near massive young stars or supernova remnants. Previous observations of low mass young stellar objects (YSOs) using ground-based observatories or the ISO-SWS instrument \citep[][]{degraauw96,vdancker_thesis} either lacked the sensitivity to observe the emission lines or the spatial resolution to separate the various emitting regions \citep[see][for an overview]{vandishoeck04}. The sensitive InfraRed Spectrograph (IRS) \citep[][]{houck04} on board the \spitzer{} Space Telescope \citep[][]{werner04} brings the detection of the mid-IR \htwo{} pure rotational and atomic fine structure lines within reach for young Sun-like stars in nearby star-forming regions. The combination of high sensitivity, moderate spectral resolution $R=\lambda/\Delta\lambda=600$, and modest spatial resolution makes \spitzer{} well suited for the study of the gas in the ISM around low-mass young stars in the nearest ($\lesssim300$\,pc) clouds. Moreover, the \spitzer{} aperture allows us, using an optimal extraction method, to disentangle both spatial components from the IRS echelle and long-slit spectral images down to scales of a few arcsec, better than can be achieved with single-dish millimeter telescopes. The combination of line fluxes, line ratios and spatial extent can help to identify the dominant heating processes -- PDR or shock-driven -- taking place, and thus the physical structure of the envelope on scales of a few hundred to a few thousand AU. The \spitzer{} data thus complement traditional studies with (sub)millimeter telescopes.

Theory predicts that disks are established early in the embedded phase of star formation, as the inevitable byproduct of the collapse of a rotating core \citep[e.g.][]{terebey84, cassen81}. Such 'young' disks, some of which show Keplerian velocity patterns, have been detected in submillimeter interferometry data \citep[e.g.][]{keene90,lommen08,joergensen09}, but little is known about their physical and chemical structure. Detailed models of X-ray and UV irradiation of disk surface layers exist for more evolved pre-main sequence stars where the envelopes have been completely dissipated \citep{bergin07,gorti08,woitke09,bergin09}. \spitzer{} observations have revealed surprisingly rich spectra of such protoplanetary disks, with the lines seen either in absorption in a near edge-on geometry \citep{lahuis06a} or in emission \citep{carr08,salyk08}. Lines from \water, HCN, \acetylene, \cotwo{} and OH have been detected, indicating hot molecular gas at temperatures of $\sim500-1000$\,K arising from the inner few AU of the disk. This hot gas is also seen in ground based CO infrared data \citep[e.g.][]{najita03, blake04}. Such disks show various atomic transitions as well, most prominently the \neii{} $12.8\mu$m line \citep{lahuis07c,pascucci06,espaillat07} \citep[see also models by][]{meijerink08}. An interesting question is therefore whether young disks in the embedded phase studied here have similar spectra as the more evolved disks, or whether they show additional features, for example due to the accretion shock of material falling on to the disk \citep{pontoppidan02,watson07}.

We present here an overview of the mid-IR gas-phase lines detected in embedded sources, including class 0 and class I sources observed in the \spitzer{} legacy program ``From Molecular Cores to Planet Forming Disks'' (``Cores to Disks'' or c2d) \citep[][]{evans03}. This program has collected a large sample of IRS spectra toward sources in the nearby Chamaeleon, Lupus, Perseus, Ophiuchus, and Serpens star-forming regions. High-$S/N$ spectra have been obtained within the $5-38\,\mu$m range for 226 sources at all phases of star and planet formation up to ages of $\sim$5\,Myr, of which 61 are presented here. Many of the sources are also included in the series of papers discussing the ice features in the c2d sample \citep{boogert08,pontoppidan08,oberg08}. In Sections \ref{sec:observations} and \ref{sec:data-reduction} the source selection and data reduction are explained. In Section \ref{sec:results} the observed atomic fine-structure and \htwo\ emission lines and the derived parameters are presented. Section \ref{sec:discussion} reviews the results within the context of PDR and shock heating and by comparing the embedded sources in the present sample with the more evolved disk sources in the larger c2d program that have been discussed in \citep[][]{lahuis07c}.

\input f1.tex

\section{Observations}
\label{sec:observations}

The data presented in this study were selected from the sample of IRS spectra observed within the \spitzer{} c2d legacy program. The c2d IRS component consists of two programs of comparable size, referred to as the first- and second-look programs. The first-look program ({PID\,\#172}) was restricted primarily to known low-mass young stars, embedded YSOs and pre-main-sequence stars with disks with stellar masses $M<2$~M$_{\sun}$ and ages $\lesssim5$\,Myr, and a sample of background stars. A few Herbig Ae stars were included as well. The c2d source selection criteria were defined to be complementary to those of the \spitzer{} legacy program ``The Formation and Evolution of Planetary Systems'' \citep[FEPS,][]{meyer07}. The second-look program ({PID\,\#179}) was, for the most part, devoted to IRS follow-up spectroscopy of sources discovered in the IRAC and MIPS mapping surveys, including a newly discovered cluster of young stars in Serpens \citep[][]{harvey06}. For all first-look observations, the integration times for the short-high (SH) and long-high (LH) modules ($R=600,\,10-37\mu$m) were chosen such that theoretical signal to noise ratios ($S/N$s) of at least 100 and 50 on the continuum were obtained for sources brighter and fainter than 500\,mJy, respectively. Deeper integrations were not feasible within the c2d program. Spectra taken using the short-low (SL) or long-low (LL) modules ($R=60-120$, 5-14 $\mu$m and 14-38 $\mu$m respectively) always reach theoretical $S/N$ ratios greater than 100. For the second-look IRS targets similar $S/N$ limits were obtained wherever possible. However, since the second-look contained a number of very weak sources (down to a few mJy) this was not always achieved.

The \nsources{} sources presented in this paper were all selected from those showing the $10\,\mu$m silicate band in absorption. This criterion includes embedded Stage I sources \citep[as defined by][]{robitaille06} plus obscured disks \citep[e.g. edge-on stage II sources ($i\gtrsim65$ degrees),][]{crapsi08}, such as CRBR\,2422.8-3423 \citep[][]{pontoppidan05a} and IRS\,46 \citep[][]{lahuis06a}. All of the sources have a rising SED in the mid-IR, the characteristic for Class I YSOs.  The selected sources are listed in Table \ref{embtab:sourcelist} which gives the basic observation and source parameters plus the adopted distances. Note that in most cases, it is not possible to determine whether the infrared source is dominated by an envelope or a disk without additional spatially resolved far-infrared and/or (sub)millimeter data. A subsequent submillimeter line study by \citet{vankempen09a} of Ophiuchus L1688 has been able to separate truly embedded YSOs (Stage I) from obscured disks or confused sources. Of the \nsources{} sources, \nemb{} are found to be truly embedded; \nedge{} sources are obscured stage II sources (edge-on disks or disks behind large columns of gas and dust); and for \nunknown{} sources the classification is unclear. Some obscured disks were not included in the c2d IRS survey of gas-phase lines in disks (\citep[][]{lahuis07c}, so their data are presented here for completeness, but are analyzed separately.

\section{Data reduction}
\label{sec:data-reduction}
The c2d reduction pipeline \citep[Chapter 3 of][and the c2d-IRS delivery documentation\footnote{http://data.spitzer.caltech.edu/popular/c2d/\\20061201\_enhanced\_v1/Documents/C2D\_Spectr\_Expl\_Supp.pdf}]{lahuis07b} was used to reduce the IRS data, starting from the S15, S16 and S17 archive data. The same c2d pipeline products as included in the final c2d Legacy data delivery\footnote{The c2d legacy data are accessible at\\{{http://ssc.spitzer.caltech.edu/legacy/c2dhistory.html}.}} were used for the spectral line analysis (see Sec. \ref{sec:analysis}). Most of the analysis focused on the SH and LH data, since for intrinsically narrow lines the SL and LL data are generally limited by the low line/continuum ratio.  The SL data were included in the analysis to search for higher-excitation H$_2\,v=0$ lines, in particular the S(3) transition.

If a module was not included in the c2d observation the \spitzer{} archive was searched for additional pointed observations to supplement the missing data. For some sources data are missing, in particular SL. For example, Serp-SMM3 and Serp-SSM4 have not been observed in the low resolution staring mode (see Table \ref{embtab:sourcelist}) and therefore do not cover the higher rotational ($J\geq3$) H$_2$ lines. Some of these are covered through mapping observations, however these are not included in this analysis.

The c2d disk sources \citep{lahuis07c} are used for comparison with the embedded sources and have all been reprocessed using the same reduction to provide the data presented in this paper. Compared with \citet{lahuis07c} this has led to an increased number of detection for the disk sources, most importantly new detections of \neii{} \citep{guedel10} and \feii{}.
%(see Appendix A) emission lines.

\subsection{Separating disk and cloud emission -- optimal extraction}
\label{sec:optimal-extraction}
A major concern when studying YSOs in the mid-IR is the possible contribution of extended (envelope or local cloud), yet structured, emission in the sometimes complex star-forming regions. Variations in the spatial distribution of the emission (see Figure \ref{fig:VSSG1-minimap}), both in the continuum and in spectral lines, limit the use of `sky' observations to correct for extended emission components. For this reason the c2d team developed an optimal extraction algorithm for IRS pointed observations starting from the SSC Archive droop products.

%GAB - Cite Evans et al. PASP article for assumed distances, especially with the possibly significant revision to the distance of Serpens?
The \spitzer{} slit size is $\sim4-5\arcsec$ for the SH module (10-19.5\,\micron) and $\sim7-10\arcsec$ for the LH module (19-37\,\micron).  At a distance of 100 pc this corresponds to physical sizes of \hbox{$\sim400-500$\,AU} and \hbox{$\sim700-1000$\,AU}, respectively.  The clouds observed in the c2d program are assumed to be located at distances ranging from 125 pc (Ophiuchus) to 260 pc (Serpens) increasing the physical area observed.  The full IRS aperture in the SH and LH spatial direction is $\sim2.5-3$ times larger than the beam size. At the observed cloud distances this means that the aperture probes physical scales of several thousand AU.  This makes it ideally suited for detecting warm or shocked H$_2$ emission from the extended (remnant) envelope, outflows, or the diffuse local cloud emission. For the sources studied in this work however the local cloud emission will potentially confuse compact disk emission.  Distinguishing between compact (disk and inner envelope) and extended (remnant envelope, outflow, or diffuse cloud) emission is therefore of vital importance for studying the emission lines originating in the circumstellar disks or the cores of jets/winds launched on small scales.

The optimal extraction developed by the c2d team uses an analytical cross-dispersion point spread function (PSF) to describe the source profile (see Figure \ref{fig:xdisp-profile}). The PSF plus an extended emission component are fit to the observed crosstalk or straylight corrected echelle (SH and LH) and longslit (SL and LL) images. The PSF is described by a sinc function with a harmonic distortion component (resulting in a significant broadening of the profile wings as shown in Figure \ref{fig:xdisp-profile}). The wavelength dependence of the PSF parameters (the order trace, the width, and the harmonic distortion) are characterized using a suite of high-$S/N$ calibrator stars. The extended emission component is approximated by a low order polynomial folded with the flatfield cross-dispersion profile. The flux calibration is derived from the calibrator stars using Cohen templates and MARCS models \citep[][]{decin04} provided through the \spitzer{} Science Center. More details about the characterization and calibration of the c2d optimal extraction is presented in Chapter 3 of \citet{lahuis07b} and in \citet{lahuis07c}. The optimal extraction returns the total flux (the source flux plus extended emission in the IRS beam) along with an estimate of the extended emission component with error estimates. The $S/N$ of the extended emission component can vary significantly depending on the quality of the raw image data and on deviations of the extended emission from the assumed uniformity across the IRS slit.

Source sizes as listed in Table \ref{embtab:features} are determined from the width of the observed cross dispersion source profile (see \hbox{Figure \ref{fig:xdisp-profile}}) deconvolved with the average width of the standard star source profiles.

\input f2.tex

\subsection{SH mini maps}
\label{sec:minimaps}
In an early phase of the c2d project \htwo\ and \neii{} lines were detected toward some of the c2d sources. Initial results using an early version of the c2d optimal extraction, from IRS starting observations, suggested some of the observed emission to be extended. Therefore, as part of the second-look program, follow-up mini-maps were taken using the SH module to check for extended emission at offsets positions of $\sim$10-15$''$ with respect to the sources. Five maps were defined to include off-source observations around eight sources. Figure \ref{fig:VSSG1-minimap} shows the results toward VSSG1, whereas \citet{lahuis07c} (their Figure 1) show the results toward the disk sources Sz\,102 (Krautter's star), Haro\,1-17, and EC\,74. Here, Figure \ref{fig:VSSG1-minimap} presents the observed \htwo\,0-0\,S(1), S(2), and \neii{} emission from the first-look on-source observations together with the off-source observations from the second-look mini maps. It is seen that most of the \htwo{} emission is extended and that it may vary on scales of a \hbox{few $\arcsec$}, especially for the S(1) line. However, some emission lines such as \neii{} are typically seen to be limited to the source itself \citep{lahuis07c}.

The mini map results confirmed the results from the c2d optimal extraction procedure. It showed that it was possible to reliable extract the extended emission component from the IRS staring observations. The procedure was further developed and has since been used in several studies, e.g. \citet{lahuis07c,geers06,boersma08}.

\subsection{1-D spectra}
\label{sec:1d-spectra}

After extraction, the 1-D spectra are corrected for instrumental fringe residuals \citep{lahuis03irs}, order matching (by a few \%) is applied, and a pointing flux-loss correction is performed on the compact source component \citep[see Chapter 3 of][]{lahuis07b}. Pointing offsets up to a few arcsec can have a noticeable impact on the derived fluxes of lines observed with the SH and SL modules, e.g. \htwo\,0-0\,S(1), S(2), \neii{}, and \neiii. For example, dispersion offsets within the nominal $3\sigma$ pointing uncertainty of \spitzer{} ($\sim$1\arcsec \ for medium accuracy peakup) can lead to SL and SH wavelength dependent flux losses up to $\sim$10\%. Fortunately for all targets a combination of modules is available which permits us to apply a reliable pointing flux loss correction.

\input f3.tex

\subsection{Spectral analysis}
\label{sec:analysis}
The IRS SH and LH modules cover the positions of the three lowest pure rotational lines of \htwo{}, fine structure emission from \neii{} (12.8 \micron), \neiii{} (15.55 \micron), \fei{} (24 \micron), \feii{} (17.9 and 26.0 \micron), \si{} (25.25 \micron), \siii{} (18.7 \micron), \silii{} (34.8 \micron), and hundreds of \water{} transitions at a resolving power of $R=\lambda/\Delta\lambda = 600$. The IRS SL module covers the higher rotation \htwo{} lines up to $J_l=7$ at a resolving power of $60-120$. Line fitting and flux integration is done to all \htwo{} and atomic lines using routines from \osia\footnote{\osia\ is a joint development of the ISO--SWS consortium. Contributing institutes are SRON, MPE, KUL and the ESA Astrophysics Division. {{http://sws.ster.kuleuven.ac.be/osia/}}}.

The extended spectral component (either the estimate or a Gaussian fit depending on the final $S/N$ in the corrected spectrum) is subtracted from the total spectrum to isolate the spatially unresolved component. Spectral lines are fit to obtain line fluxes for both spatial components. Uncertainty estimates, as listed in Tables \ref{atomic:tab:linefluxes} and \ref{h2:tab:linefluxes}, are derived from the residuals after line fitting, or, in the absence of a spectral line, using the line width derived from the instrumental resolution. An uncertainty estimate is derived for both the extended and compact line fit and combined. As a result, the 1-$\sigma$ uncertainty estimates can vary widely for sources with a similar continuum flux and integration time. This may, for example, be the result of the presence of artifacts resulting from hot pixels or small variations in the extended emission which are not fully accounted for in the spectral extraction.

Typical mean 4$\sigma$ uncertainties prior to subtraction of the extended component for the high resolution modules range from $\sim$0.1-2$\times10^{-15}\ \mathrm{erg\,cm^{-2}\,s^{-1}}$ with extremes of $\sim$5$\times10^{-17}\,\mathrm{erg\,cm^{-2}\,s^{-1}}$ and $\sim$1$\times10^{-14}\,\mathrm{erg\,cm^{-2}\,s^{-1}}$. Note that the uncertainties used in this paper include an extended component uncertainty.

\section{Results}
\label{sec:results}

A search for the major atomic fine-structure lines and \htwo{} lines in the \spitzer-IRS wavelength range has been performed for all \nsources{} sources in our sample. Line detections are shown in Figures \ref{fig:atomic_lines} and \ref{fig:H2_lines} for the embedded sources and described in Sections \ref{sec:lines_analysis} and \ref{sec:h2_analysis}. In Figure \ref{fig:line_relations} the measured line strengths and upper limits, converted to solar luminosities, are shown as function of $L_{\mathrm IR}$ (the continuum luminosity between $\lambda=12.8-15.5\,\mu$m), see Section \ref{sec:discussion-correlations}. Figure \ref{fig:line_ratios} shows the ratios and ratio limits for a number of lines. Lines of molecules \water, \acetylene, HCN and \cotwo{} have also been searched for and are discussed in Section \ref{sec:molecules}.

For both the atomic lines and those of \htwo\ the observed upper limits are in most cases of the same magnitude as the median line fluxes of the detections (see Figure \ref{fig:line_relations}). It is therefore likely that for most sources emission in one or more lines will be present but has gone undetected. The upward trend observed in the upper limits above an IR luminosity of $\sim10^{-3}\mathrm{L}_\odot$ is the result of the dominance of target shot noise in the IRS detector noise at higher source signals.

\subsection{Atomic fine-structure lines}
\label{sec:lines_analysis}

Figure \ref{fig:atomic_lines} shows all 4$\sigma$ detections of atomic lines. One or more of the \neii{}, \feii{} 18 and 26\,\micron,  \si, and \silii{} emission lines are observed toward $\sim$2/3 of the \nemb{} embedded sources. The measured line fluxes of the extended and compact components and the upper limits are listed in Table \ref{atomic:tab:linefluxes}. Of the sources with line detections $\sim$30\% show emission from more than one line. Emission of most atomic lines is observed as both compact and extended emission. 

\input f4.tex

\subsection{Molecular hydrogen}
\label{sec:h2_analysis}
Figure \ref{fig:H2_lines} shows all 4$\sigma$ \htwo{} emission line detections toward the 43 embedded sources in our sample. Emission from one or more \htwo{} lines observable with the \spitzer-IRS (S(0) to S(7)), has been detected toward 36 out of 43 sources, with 31 out of 43 showing low$-J$ lines and 19 out of 43 high$-J$ lines. The \htwo{} line fluxes for the compact and extended components and the line flux upper limits are listed in Table \ref{h2:tab:linefluxes}. The warm gas traced by the low$-J$ lines is observed primarily in the extended component while the hot gas probed by the high$-J$ lines is observed mostly in the compact component.

The warm gas (\tex$\lesssim700$\,K) traced by the lower rotational lines ($J_l\leq2$) is mostly extended, except for a few cases. Hot (\tex$\sim1000-1500$\,K) gas is traced by the higher rotational lines ($J_l\geq4$). For $J_l\geq 5$ no estimate of extended emission is available because the IRS is spatially undersampled at these wavelengths and the usability of the optimal extraction is limited. Mapping observations \citep[e.g.][]{neufeld09} show that, in general, the \htwo{} high-$J$ lines have a similar distribution to the lower excitation lines. Since most of the $J_l=4$ detections show no strong signs of an extended emission component, we assume all hot \htwo{} emission to have a compact origin.

\input f5-6.tex

In the simplest analysis, the \htwo{} excitation is assumed to be in local thermal equilibrium (LTE) \citep[e.g.,][]{thi01} with an ortho-to-para ratio determined by the kinetic temperature of the gas \citep*[following][]{sternberg99}. For gas temperatures 100, 150, and $\geq$200\,K, the ortho-to-para ratios are 1.6, 2.5, and 3, respectively. Assuming optically thin emission, the integrated flux of a rotational line $J_u \rightarrow J_l$ for a given temperature \tex{} is

\begin{equation}
   \label{eq:F}
   F_{ul}(T_\mathrm{ex}) = \frac{hc}{4\pi\lambda}
            N(\mathrm{H}_2) A_{ul} x_u(T_\mathrm{ex}) \Omega
            \ \mathrm{erg\,s^{-1}\,cm^{-2}},
\end{equation}
where $\lambda$ is the wavelength of the transition, $N(\mathrm{H}_2)$ the total column density, $A_{ul}$ the spontaneous transition probability, and $\Omega$ the source size. For high enough densities ($n>10^3\mathrm{cm}^{-3}$), the population $x_u$ follows the Boltzmann law
\begin{equation}
   \label{eq:x}
   x_u(T_\mathrm{ex}) = \frac{g_\mathrm{N}(2J_u+1) e^{-E_J/kT_{\mathrm{ex}}}}
              {Q(T_{\mathrm{ex}})}
\end{equation}
where $E_J$ is the energy of the upper level, $g_\mathrm{N}$ is the nuclear statistical weight (1 for para and 3 for ortho \htwo), and $Q$(\tex) the partition function for the given excitation temperature \tex.

Using the above equations, excitation temperatures, column densities and \htwo{} gas masses, or limits on these, can be derived from the observed line fluxes and upper limits. A source size has to be known or assumed to derive column densities. Observations of multiple \htwo{} lines in PDRs and shocks typically show multiple temperature components \citep{vdancker99,habart05}, e.g. a warm ($T_\mathrm{ex}$ of a few 100\,K) and a hot ($T_\mathrm{ex}\sim1000$\,K or higher) component. In our data, often only a few lines per source are detected. If $J_l<2$ lines are detected, a (limit of the) temperature of the warm gas is derived. For a few sources more than one of these lines are detected in the extended component. The derived temperatures vary from 85 to 120\,K with uncertainties of $\sim 20$\,K. Therefore we adopt a temperature of 100\,K for the warm component for all sources. If two or more higher excitation lines ($J_l>2$ and higher) are detected a temperature for the hot component is derived; otherwise a temperature of 1000\,K is adopted. This is done for both spatial components. For some sources multiple H$_2$ lines are detected allowing a temperature estimate to be made. Temperatures for the hot component, varying from $\sim500-1500$\,K, are observed (see Figure \ref{fig:ex_diagram}) in both spatially extended and compact emission. In the analysis no extinction correction is applied and only \htwo{} is included in the mass calculation.

The column density averaged over the IRS aperture is derived using above equations. For all compact source emission the source size is smaller than the IRS aperture (Sec. \ref{sec:data-reduction}). Since this is unknown, a typical size $r=50$\,AU is adopted (gas masses up to a few $10^{-3}$\,Jupiter masses, see below and Figure \ref{fig:h2_mass}, give an upper limit to the emitting area of $50$\,AU assuming a density $n\gtrsim10^5\,\mathrm{cm}^{-3}$). The fitted or adopted excitation temperature plus the $J_\mathrm{up}$ column densities provide the total column density for that temperature, which in turn gives the total H$_2$ gas mass,
$M_{\mathrm{H}_2} = { \pi r^2 \times N \times 2 m_\mathrm{H}} / M_\mathrm{J}$
with $m_\mathrm{H}=1.674\cdot 10^{-24}$\,gr and $M_\mathrm{J}=1.9\cdot 10^{30}$\,gr, excluding a mass contribution by He. Note that the derived column density scales as $1/r^2$ while the derived gas mass is independent of the adopted beam and/or source size. Therefore the derived column densities can only be used as order of magnitude estimates and are listed for both the warm and hot gas in Table \ref{embtab:gasparam}. Figure \ref{fig:h2_mass} shows the \htwo{} mass distribution of the warm (\tex\,$=100$\,K) and hot (\tex\,$=1000$\,K) components for the truly embedded sources. H$_2$ masses of a few Jovian masses for the warm gas and a few milli-Jovian masses for the hot gas are observed. Hatched bars include all sources where \htwo{} is detected.

\input f7.tex
\input f8.tex

\subsection{Molecular emission}
\label{sec:molecules}

Features of \water{}, \acetylene, HCN and \cotwo{} emission are searched for in the full sample. Detecting these species is difficult because the emission features are generally weak and ride on top of a strong mid-IR continuum. \water{} emission is most easily detected as the spectra contain a large number of both spectrally unresolved and blended emission lines across the complete IRS-SH spectral range. Thus, a full spectral range correlation can be applied. Detecting \acetylene, HCN and \cotwo{} is more difficult because only the Q-branches can be observed (P- and R-branch lines are undetectable at the IRS resolution). The Q-branches are, moreover, weak and partially blended with \water{} lines, as seen for example in the spectra of VSSG17 and L1689\,IRS5 spectra in Figure \ref{fig:water} where the \acetylene, HCN and \cotwo{} Q-branch positions are marked.

Toward 12/43 truly embedded sources spatially unresolved \water{} emission is positively identified and the presence of \acetylene, HCN and \cotwo{} is, by visual inspection, observed toward a small number of sources (see, e.g., the spectrum of VSSG17 in Figure \ref{fig:water}). Toward 4 sources \water{} is clearly detected in both the optimal extraction and the full aperture extraction spectra. For 11 more sources \water{} is positively detected in the optimal extraction spectrum and tentatively detected in the full aperture spectrum. Figure \ref{fig:water} shows the continuum subtracted SH spectra with a LTE water model overplotted for two of the embedded sources. The observed emission features are similar to those seen in the more evolved disks around T\,Tauri stars by \citet{salyk08,carr08}.

To derive basic excitation parameters the full IRS SH spectra are compared with single temperature LTE models using the HITRAN 2004 database \citep{rothman05}. The model assumes a single slab of gas of solid angle ($\Omega$), temperature ($T$), column density ($N$) and intrinsic linewidth ($b$).  Without exception the derived excitation temperatures are high, \hbox{\tex\,$=700\sim1500$\,K}. Best fits are achieved for models with large \water{} column densities, $N\gtrsim10^{17}\,\mathrm{cm}^{-2}$, and small emission regions, $r\lesssim$\,few\,AU, comparable to what is found for the T\,Tauri stars observed by \citet{salyk08,carr08}. An accurate and fully constrained fit is not possible with the low ($R=600$) resolution SH spectra. It may be possible to extract more information through non-LTE models \citep[see][]{meijerink09} but such models are outside the scope of this paper.

For \acetylene, HCN and \cotwo{} none of the spectra have sufficient $S/N$ to derive reliable excitation temperatures and column densities or even solid detections. High spatial and/or spectral resolution observations are required to further constrain the model parameters. The importance of high resolution data is illustrated in \citet{salyk08} who obtained high resolution $3\mu$m spectra for two disk sources, AS\,205A and DR\,Tau, following the initial detection of hot \water{} emission in the \spitzer-IRS SH spectra. We therefore do not list fit parameters for individual sources but limit ourselves to summarizing detections in Table \ref {embtab:features}.

\subsection{Correlations and line ratios}
\label{sec:discussion-correlations}

Figure \ref{fig:line_relations} shows the observed line luminosities (for all 4$\sigma$ detections) as function of the mid-IR ($12.8-15.5$\,\micron) continuum luminosities as calculated from the \spitzer-IRS spectra. The mid-IR continuum luminosity is used as a qualitative measure of the hot-dust component.

Both the line strengths and the upper limits fall in a narrow band, $\sim$2 orders of magnitude, spanning the full $L_\mathrm{IR}$ range. All species show a positive trend with $L_\mathrm{IR}$, which is most clearly observed in the compact \htwo\,S(4) and \neii\ line emission. As discussed above, the presence of weaker line emission than currently observed cannot be excluded. However, the current dataset does exclude strong line emission for the weak IR sources.

Figure \ref{fig:line_ratios} shows the line ratios and line ratio limits of \neii\ with a number of \htwo{} and atomic lines. All ratios fall within a range of a few orders of magnitudes and show no obvious trends. The line ratio limits do not reveal any sources deviating from this, consistent with the fact that the observed line luminosities are close to the line luminosity upper limits as shown in Figure \ref{fig:line_relations}. Higher resolution and high sensitivity observations (e.g. with the MIRI instrument aboard the James Webb Space Telescope) will be required to overcome this limitation.

%\textbf{Which extra ratios would be usefull?} 

\input f9.tex

\section{Discussion}
\label{sec:discussion}

\subsection{General trends}
Atomic fine structure lines, \htwo{} pure rotational lines and molecular emission features are observed toward the majority of the sources in our sample, both as compact and extended emission. Table \ref{embtab:features} presents a qualitative overview of the identified emission. The compact emission is dominated by \neii{}, high-$J$ \htwo{} emission lines and hot \water{} emission. The extended emission is dominated by low-$J$ \htwo{} emission lines. A small number of sources show line emission from both the extended and compact components at the same time. \water{} emission is seen in the compact component toward both compact and extended emission dominated sources.

The increase of the line strength with increasing mid-IR luminosity suggests that the heating mechanism of the dust and the excitation of the line may have the same physical origin. Besides heating due to the protostellar luminosity, UV radiation from the star-disk accretion boundary layer escaping through the outflow cones may heat both the dust and the gas (see \S 5.3).

A global comparison in Figures \ref{fig:line_relations} and \ref{fig:line_ratios} between the truly embedded sources and the disk sample of \citep{lahuis07c} shows that both samples exhibit spatially compact and extended emission at the same time. Both samples also show compact hot water emission. The main difference is in the statistics of the extended emission. While in the compact emission the difference between the two samples is small (e.g. hot water emission is observed toward $\sim$40\,\% of the disk sources and $\sim$30\,\% of the truly embedded sources) the difference is large for the extended component. Extended emission is observed toward $\sim$1/3 of the disk sources and $\sim$2/3 of the truly embedded sources. The extended emission observed toward the disk sources represents the PDR emission from the local environment, in particular the extended PDR in $\rho$\,Oph. Therefore it is fair to assume that a significant fraction of the extended emission observed toward the truly embedded sources has its origin from the source and not from the environment. 
%***this statement may be biased by Ophiuchus ***

\input f10.tex

Because the \spitzer-IRS spectral and spatial resolution is moderate, the level to which we can qualitatively and quantitatively constrain the excitation and the physical environment of young stars is limited. High spatial and/or high spectral resolution observations are required to separate the different emission components either through direct observations or by spectral decomposition. A prime example of this is the study of \neii{} emission toward stage II disk sources. The initial \spitzer{} detections of compact \neii{} emission \citep[][]{lahuis07c,pascucci07} raised excitement and suggested a possible disk surface origin of the \neii{} emission which was supported by model results from \citet{Glassgold07}. However, recent high resolution ground based observations \citep[][]{herczeg07,pascucci09,vanboekel09} have shown that high resolution follow-up observations are essential for a proper interpretation. \citet{pascucci09} and \citet{vanboekel09} argue for an outflow/jet origin of the spatially unresolved \spitzer{} \neii{} emission observed toward T\,Tau, Sz102 and VW\,Cha. Recent model results \citep{hollenbach09} confirm that observable \neii{} emission may be produced through X-ray and FUV irradiation of the proto-planetary disk surface and through internal shocks in protostellar winds. This is confirmed by \citet{guedel10} where disk and jet contributions are separated through a statistical analysis of \neii{} emission of a large sample of \spitzer{} sources.

Although a detailed analysis of individual sources may not be possible, general conclusions may be drawn. By looking at the overall line fluxes and line ratios observed \hbox{(see Figures \ref{fig:line_relations} and \ref{fig:line_ratios})} we can draw general conclusions about the protostellar environment. In the following sections we will therefore compare the observations with published shock and PDR models and discuss possible origins of the observed emission.

\subsection{Shock excitation}
\label{sec:shocks}

Many of the observed emission lines, including \neii, \feii, \si{} and \silii{} are predicted to be strong in $J$-type shocks \citep{hollenbach89} with the main exception being the \hbox{low-$J$} \htwo{} lines. The hot ($\sim$1500\,K) \htwo{} emission observed, in particular, is expected in the cooling post-shock gas \citep[e.g.][]{vdancker99}. The observed strengths of the atomic emission lines are comparable with predictions by shock models, but the observed line ratios are not sufficiently constrained to draw any conclusions about precise shock velocities. However for \neii{} to be strong, shock velocities in excess of $\sim70$\,km\,s$^{-1}$ are required. Figure \ref{fig:line_ratios_shock} shows that the range of line ratios predicted by the models is consistent with those observed (Figure \ref{fig:line_ratios}) over a range of expected pre-shock densities and shock velocities.

The \htwo\,S(0) line is predicted to be the weakest \jshock{} feature in \citet{hollenbach89}. The observed line ratios and upper limits of the \neii/\htwo\,S(0) line ratio for the extended component in the disk sources \citep{lahuis07c} is $\leq$1 (see Figure \ref{fig:line_ratios}). In \jshock{} models the \neii/\htwo\,S(0) ratio is up to two orders of magnitude higher (see Figure \ref{fig:line_ratios_shock}). We therefore exclude \jshocks{} as the origin for the extended low-$J$ warm \htwo\ emission toward the embedded and disk sources. The extended \feii{} emission observed toward the disk sources may have a similar origin as the warm \htwo{} emission though this cannot be constrained by the $J$-shock models itself.

In non-dissociative \cshocks{} low-$J$ \htwo{} emission is strong \citep{kaufman96} and the observed strengths of the \htwo{} lines are consistent with \cshock{} model predictions. At the same time atomic lines are not expected to be strong. The low-$J$ \htwo{} lines, mostly observed in the spatially extended component, may partly have a \cshock{} origin in material entrained by the outflow. However given that \feii{} is also observed as extended emission, \cshocks{} are not likely to be the sole origin of the extended line emission.

% critical densities
% Si+  2.6e5
% Fe+  2.2e6
% H2 S(1) 2.3e4
% H2 S(2) 2.2e5
% H2 S(3) 9.4e5

%PDR Kaufman 2006
% Si II consistent with 10^2-10^3 Go at n=10^5
% Fe II order of magnitude lower
% S(0) ~5x lower
% S(1) 10-6-10-4
% S(2)  id.
% S(3)  id.

\subsection{PDR models}
\label{sec:pdr}

Models of photodissociation regions (PDRs) predict line strengths for the low-$J$ \htwo, \feii{} and \silii{} emission comparable to the observed line strengths for $n\sim10^2-10^5\,\mathrm{cm}^{-3}$ and a radiation field of $\sim10^1-10^3$\,G$_0$ \citep{hollenbach91,kaufman06}. These values are consistent with those estimated for the outer envelope. The \htwo{} excitation temperatures of $\sim350-700$\,K (see Figure \ref{fig:ex_diagram}) are comparable with gas surface temperatures for PDRs with the range of densities and radiation fields mentioned above \citep[][and the PDRT\footnote{Photo Dissociation Region Toolbox\\\url{http://dustem.astro.umd.edu/pdrt/}}]{kaufman99}.

A PDR origin of the observed extended emission therefore appears to be plausible. It is likely that the observed emission originates in the PDR tracing the surfaces of the extended outflow cavity for the embedded sources, although there may also be a contribution from extended lower density cloud material in the beam for some sources, especially in Ophiuchus. Weak \cshocks{} may be present as well (see Section \ref{sec:shocks}). The presence of extended photon-heated gas along outflow cavities has recently also been inferred from narrow high-$J$ CO submillimeter emission lines \citep{vankempen09b,vankempen09c,spaans95} and has also been suggested based on ISO-LWS atomic fine-structure line data \citep{molinari01}.
%\textbf{, while \citet{spaans95}
%  argue for a PDR origin in the envelope surrounding the outflow
%  cavity to explain the narrow $^{12}$CO and $^{13}$CO J=6--5 emission
%  observed toward many low-mass YSO's.}
%GAB - Should the Spaans et al. CO 6-5 paper be referenced here?
%FL - yes, I think so 
Although the \si{} emission is almost exclusively observed in the extended component, \si{} is not strong in PDR models. At the warm PDR surfaces most sulfur will be in the form of S$^+$.

The origin of the exciting UV radiation field is still under debate (see \S 1). Suggested origins are UV photons: i) escaping from the star-disk boundary layer near the central protostar \citep{spaans95}, and ii) from high-velocity shocks directly associated with the more extended jets and outflows \citep{neufeld89a,molinari01}. In both cases, the UV field consists of a continuous spectrum with possibly some discrete lines due to recombining atoms superposed. In our sample we observe sources simultaneously exhibiting signatures of extended PDR/\cshock{} and compact \jshock{} emission and sources showing only extended PDR/\cshock{} emission and no compact shock emission (see Table \ref{embtab:features}), thus leaving all options open.

In order to advance on the current situation it will be important to get multiple, when possible spatially and/or spectrally resolved, detections per source on key lines such as \neii{}, \si{}, \oi{} 63.2\,$\mu$m, \cii{} 157.7\,$\mu$m and the low-$J$ H$_2$ lines. \neii{} and \si{} are clear indicators of \jshocks, whereas the \oi/\cii{} ratio and the absence or presence of \silii{} and \feii{} help to distinguish and characterize PDR and \cshock{} emission. In his PhD thesis \citet[][Chapter\,10]{vdancker_thesis} presents a useful summary of the relevant issues. High velocity resolution \neii{} observations are possible from the ground although the lines presented here are not strong enough for current instrumentation. JWST will provide an inventory of the key mid-IR emission lines towards a large sample of embedded sources. \herschel{} will provide the possibility to observe highly-excited CO, \oi{} and \cii{} and provide high velocity resolution observations for some far-IR lines through HIFI.

\subsection{Molecular emission and absorption}

While molecular emission is seen toward a number of sources, none of the truly embedded sources show absorption bands of HCN, \acetylene{} and/or \cotwo.  This in contrast with high and intermediate mass stars \citep[][Thi et al. in prep.]{lahuis00,boonman03a,boonman03b} where these molecules are characteristic of hot-core chemistry \citep{doty02,rodgers03}. Toward ULIRGs absorption of \acetylene, HCN and \cotwo{} has been observed by \citet{lahuis07a} suggesting the presence of pressure confined massive star formation. The absence of signatures of \acetylene{} and HCN absorption towards the low-mass sources in our sample is not surprising, however. To produce absorption features observable with IRS one needs both high column densities, requiring orders of magnitude enhancement of the molecular abundances compared with the normal cloud abundances, and a strong infrared background. These conditions are met in the inner envelope around higher-mass stars during the hot-core phase \citep{lahuis00}, but for solar-mass stars the inner envelope regions do not become hot enough ($T>300$\,K) to produce the required chemical enrichment. Only in the inner regions of protoplanetary disk are these conditions met and absorption features have only been observed with \spitzer{} and near-IR ground-based instruments \citep{lahuis06a,gibb07} under a favorable, near edge-on, source alignment toward two sources (IRS\,46 and GV\,Tau).

Molecular emission from \water{} is observed in a quarter of the truly embedded sources in our sample and emission from \acetylene, HCN and \cotwo{} is tentatively observed in a few sources. What is the origin of this emission: the young disk or outflow?

Mid-infrared \water{} and OH emission is observed to be associated with outflows such as HH 211, where prominent lines are detected at the bow shock position offset from the protostar itself \citep[e.g.]{tappe08}. Emission features of \acetylene, HCN and \cotwo{} seen toward Cepheus A East are evidence of gas-phase abundance enhancements through sputtering of icy grain mantles in shocked regions \citep{sonnentrucker06,sonnentrucker07}. Since our data show neither the high-$J$ OH lines nor prominent \acetylene{}, HCN or \cotwo{} emission, these scenarios do not appear to explain our data.

Strong \water{} mid-infrared emission has also been observed with \spitzer{} toward the \hbox{stage 0} source \citep[NGC1333-IRAS4B by][]{watson07} and attributed to disk accretion shocks, but it is unclear whether this interpretation is unique and how general this phenomenon is \citep[see also][]{joergensen10}. None of the other stage 0 sources in our sample (in particular VLA\_1623-243, IRAS\_16293-2422, IRAS\_16293-2422B, Serp-S68N, Serp-SMM3 and Serp-SMM4) show hot \water{} emission. One Stage I source, GSS30-IRS1, exhibits strong CO mid-infrared lines indicative of accretion shocks \citep{pontoppidan02} but does not show any \water{} lines in our {\it Spitzer} data.

On the other hand, strong \water, \acetylene, HCN, \cotwo\ and/or OH lines have clearly been detected from a number of more evolved stage II sources arising from the inner AU of the disk \citep{carr08,salyk08}. The molecular emission observed toward the embedded sources in our sample has characteristics similar to those observed toward these disk sources: high $T_\mathrm{ex}$, high $n$ and a small size of the emitting region. In particular, $T_\mathrm{ex}$ is significantly higher than the value of $\sim$170 K observed toward NGC 1333 IRAS4B \citep{watson07}. It is therefore plausible that the molecular emission observed in our sample also has its origin in the inner AU of the young disks. In other words, we may be looking through a window onto the embedded young disks \citep[see e.g.][]{cernicharo00}.

\input f11.tex

\subsection{Source environment}
\label{sec:source_env}

The general picture which emerges is that of a young star surrounded by a forming disk embedded in the natal envelope. Winds and jets from the young star produce holes and cavities in the envelope. Within the jets and in its boundary regions with the envelope, supersonic shock waves are produced which locally compress and heat the gas producing, among others, hot \htwo{} and \neii{} emission. Non-dissociative \cshocks{} in material entrained by the extended outflow produce warm spatially extended \htwo{} emission. \cshocks{} are however not the sole origin of the observed extended emission: the presence of spatially extended \feii{} and \silii{} emission argues against this and points to the presence of extended PDRs. UV radiation produced by the young star system heat the cavity walls and envelope surface creating extended PDR regions. Emission from the young disk, in particular that from molecules such as H$_2$O, can be observed directly by looking along the outflow cavities and/or through a patchy envelope.

Spatially extended PDR and \cshock{} emission is observed toward $\sim$2/3 of the sources, including the stage 0 sources. Some of this emission may come from the local environment, as observed toward the disk sources. Extended emission is observed toward $\sim$1/3 of the disk sources. Therefore a significant fraction of the extended emission observed toward the embedded sources may originate from the sources themselves. This is consistent with most sources having a sizeable outflow cavity. The outflow cavities are most easily observed toward sources where the outflow is seen face-on or at moderate inclination angles. Toward systems seen at large inclination angles the extinction through the envelope will prohibit detecting the emission from the outflow cavity walls, though this will depend on the observed wavelength of the emission and its location within the cavity emission. Thus, given a uniform line-of-sight distribution, extended emission from the outflow cavity walls should be observed towards a large fraction of the sources. This is the case in our sample of embedded sources.

The presence of hot \water{} emission suggests a strong similarity between the embedded stage-I disks and the more evolved stage-II disks. None of the known stage-0 sources in our sample show signs of hot \water{} emission, implying that their disks have either different characteristics or are much more obscured.

Figure \ref{fig:cartoon} shows a graphical representation of the picture described above displaying the different components and the location of the observed emission. In this picture, the main difference with the more evolved stage II disks is that the latter sources lack the extended emission from PDR and C-shocks along the cavity edge. The limitations of the \spitzer-IRS data (low spatial and spectral resolution) prohibit us from making conclusive statements on individual sources.  

\section{Conclusions}

A survey of the mid-infrared gas phase pure rotational lines of molecular hydrogen and a number of atomic fine structure transitions has been carried out toward a sample of \nsources\ class I sources with the \spitzer-IRS of which \nemb{} are truly embedded protostars. Both extended and compact source emission has been detected toward a majority of the sources. Compact emission, most prominently seen in hot \htwo{}, \neii{} and \water{} lines, is seen toward $\sim$1/4 to 1/3 of the truly embedded sources. The compact hot \htwo{} and \neii{} emission is believed to have its origin predominantly in dissociative \jshocks{} in the outflow jets, though a true disk component may be present. The \water{} emission is consistent with an origin in the inner regions of the young protostellar disk, similar to the \water{} emission from the more evolved systems. Extended emission, dominated by low-$J$ \htwo{} and \feii{} emission, is observed toward $\sim$2/3 of the truly embedded sources, both stage 0 and I sources. The extended emission is thought to originate from PDRs along the outflow cavity walls with possible contributions from non-dissociative \cshocks.

These observations have shown the capability of \spitzer{} to observe the gas around young stars. It clearly shows emission from PDR and shock excited gas by the presence of various \htwo{} and atomic lines. However, its limited spatial and spectral resolution prevents a quantitative analysis of the excitation conditions.

Sources with atomic fine structure line emission, \htwo{} and \water{} emission are excellent candidates for follow up with various ground based instruments and near-IR and far-IR space-based instruments on \herschel{} and JWST. The major far-IR cooling lines of [O\,{\sc i}] and [C\,{\sc ii}] as well as high-$J$ CO observable with \herschel{} can provide important excitation constraints \citep{vankempen10}. Several of the sources in this sample are targets in the ``Water in Star-forming regions with Herschel'' (WISH) and ``Dust, Ice and Gas in Time'' (DIGIT) \herschel{} key programs. Future mid-IR observations with JWST-MIRI will allow the detection of weak emission lines and provide many more diagnostics to separate the physical processes. Such high spectral and spatial resolution observations are essential to constrain the location and extent of the emission (either by direct detection or through velocity resolved spectra) to properly understand the physical environment around the young stars.

The spectra of the compact component of the truly embedded sources show remarkable similarities with those of the evolved disk sources, both in terms of absolute fluxes and flux ratios. Both characteristic \jshock{} emission, hot \htwo{} and \neii, and true disk emission through the molecular emission lines are observed. In particular, the molecular emission may offer a unique tool to study the young forming disks in their earliest phases and compare these with the more evolved stage II disks.

\begin{acknowledgements}
Astrochemistry in Leiden is supported by a NWO Spinoza grant and a NOVA grant. Support for this work, part of the Spitzer Legacy Science Program, was provided by NASA through contract 1224608 issued by the Jet Propulsion Laboratory, California Institute of Technology, under NASA contracts 1407, 1256316, and 1230779. We thank the referee, David Hollenbach, for his thorough and constructive review which helped to improve our paper. All members of the c2d-IRS team are thanked for their feedback; detailed comments from Klaus Pontoppidan are particularly appreciated. We thank the Lorentz Center in Leiden for hosting several meetings that contributed to this paper. Lars Kristensen and Tim van Kempen helped to identify the evolutionary stages of some sources.
\end{acknowledgements}

\addcontentsline{toc}{chapter}{\bibname}
\bibliography{thesis}
\bibliographystyle{aa}

%\renewcommand{\rightmark}[1]{Tables}
%\markboth{Chapter \thechapter. \ \shortchapter}{Tables}
%\strut\vspace{15em}
%%\begin{center}
%{\Huge Tables}
%\end{center}
\clearpage
\ifodd \thepage
\parbox{\textwidth}{
\strut\vspace{5cm}
\begin{center}
{\Huge Tables}
\end{center}
}
\clearpage
\else
\fi

\renewcommand{\baselinestretch}{1.0}

\input tab1.tex
\clearpage
\input tab2.tex
\clearpage
\input tab3_1.tex
\clearpage
\input tab3_2.tex
\clearpage
\input tab4_1.tex
\clearpage
\input tab4_2.tex
\clearpage
\input tab5.tex

\end{document}

%% file: f1.tex
\begin{figure*}[t]
\begin{center}
\includegraphics[width=0.45\textwidth]{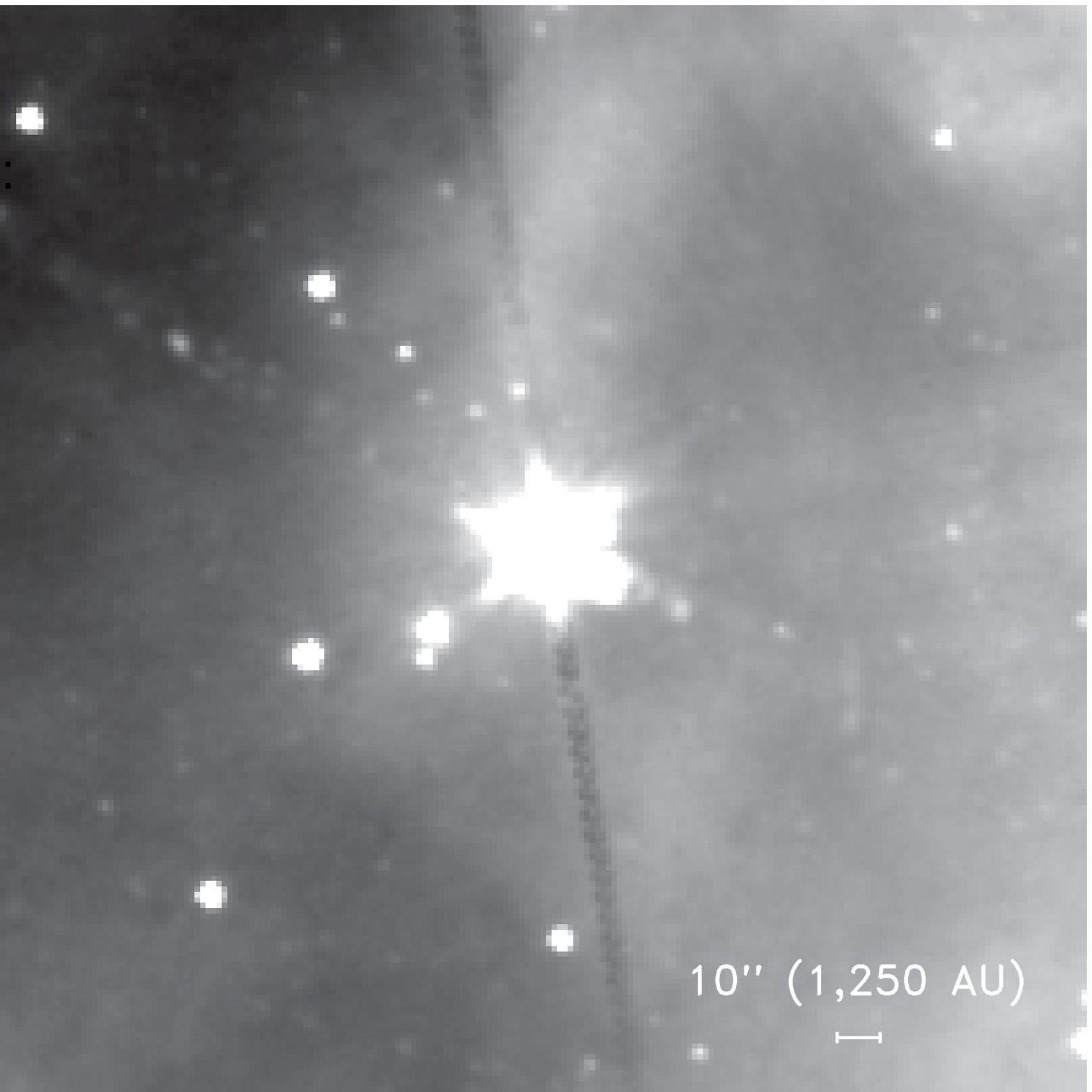}%
\includegraphics[width=0.5\textwidth]{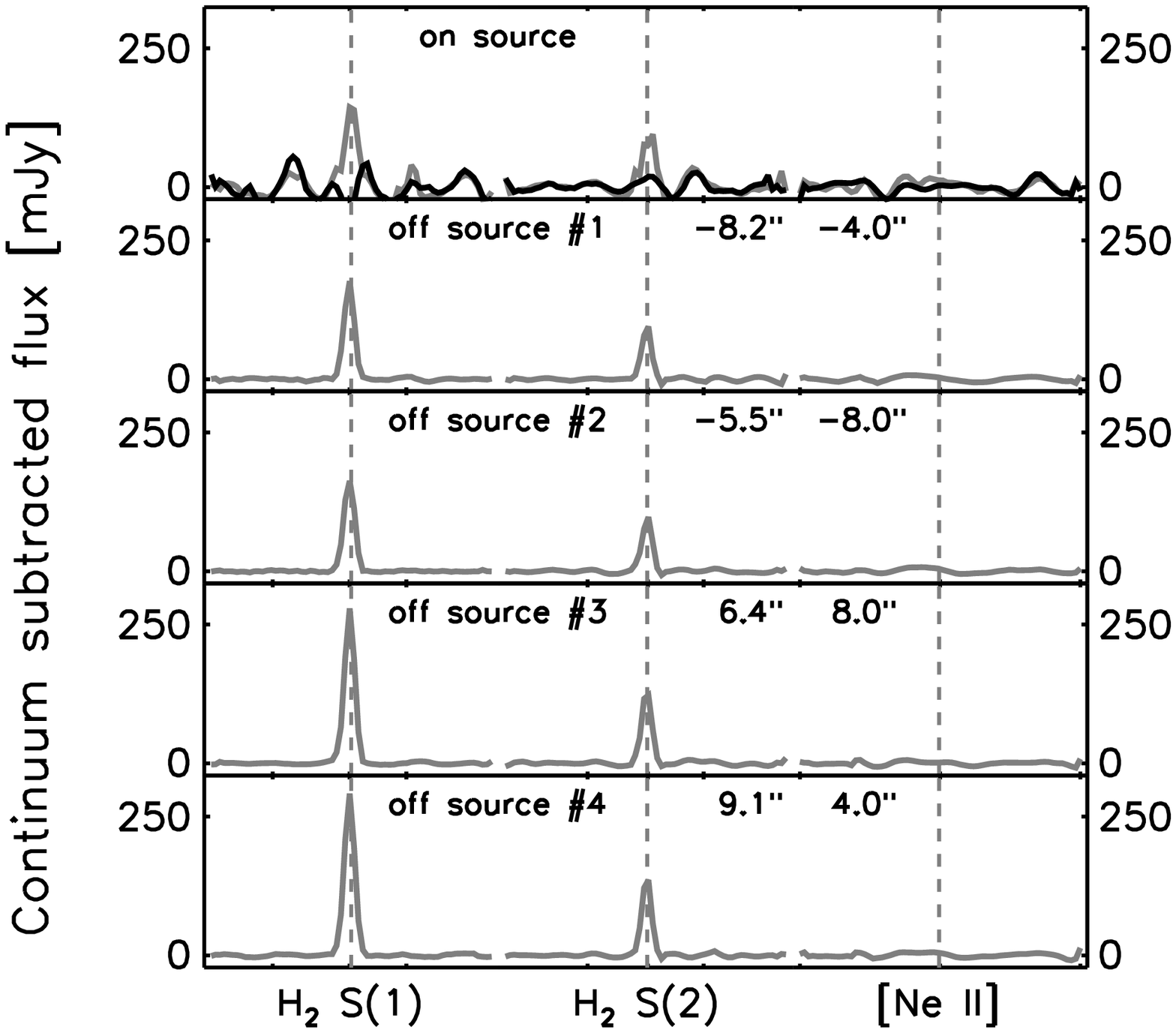}
\caption{\small\label{fig:VSSG1-minimap}%
\textit{Spitzer}-IRS SH minimap observed around VSSG1 in Ophiuchus.
The left plot displays a \textit{Spitzer}-IRAC image at 4.5 $\mu$m
(including the high excitation H$_2$ S(9), S(10) and S(11) lines)
showing the star plus diffuse extended emission. The right plot shows
the spectrum observed in the on-source \textit{Spitzer}-IRS observation
(top panel) and the off-source spectra obtained with the 
\textit{Spitzer}-IRS minimap. In gray the total observed emission and
in black the unresolved source emission toward VSSG1 are presented.
Virtually all H$_2$ S(1) 17.03 $\mu$m and S(2) 12.28 $\mu$m emission
is extended in this source. \neii{} 12.81 $\mu$m emission is absent 
toward VSSG1.
}
\end{center}
\end{figure*}

%% file: f2.tex
\begin{figure*}[t]
\begin{center}
\parbox{0.55\textwidth}{\includegraphics[angle=90,width=0.55\textwidth]{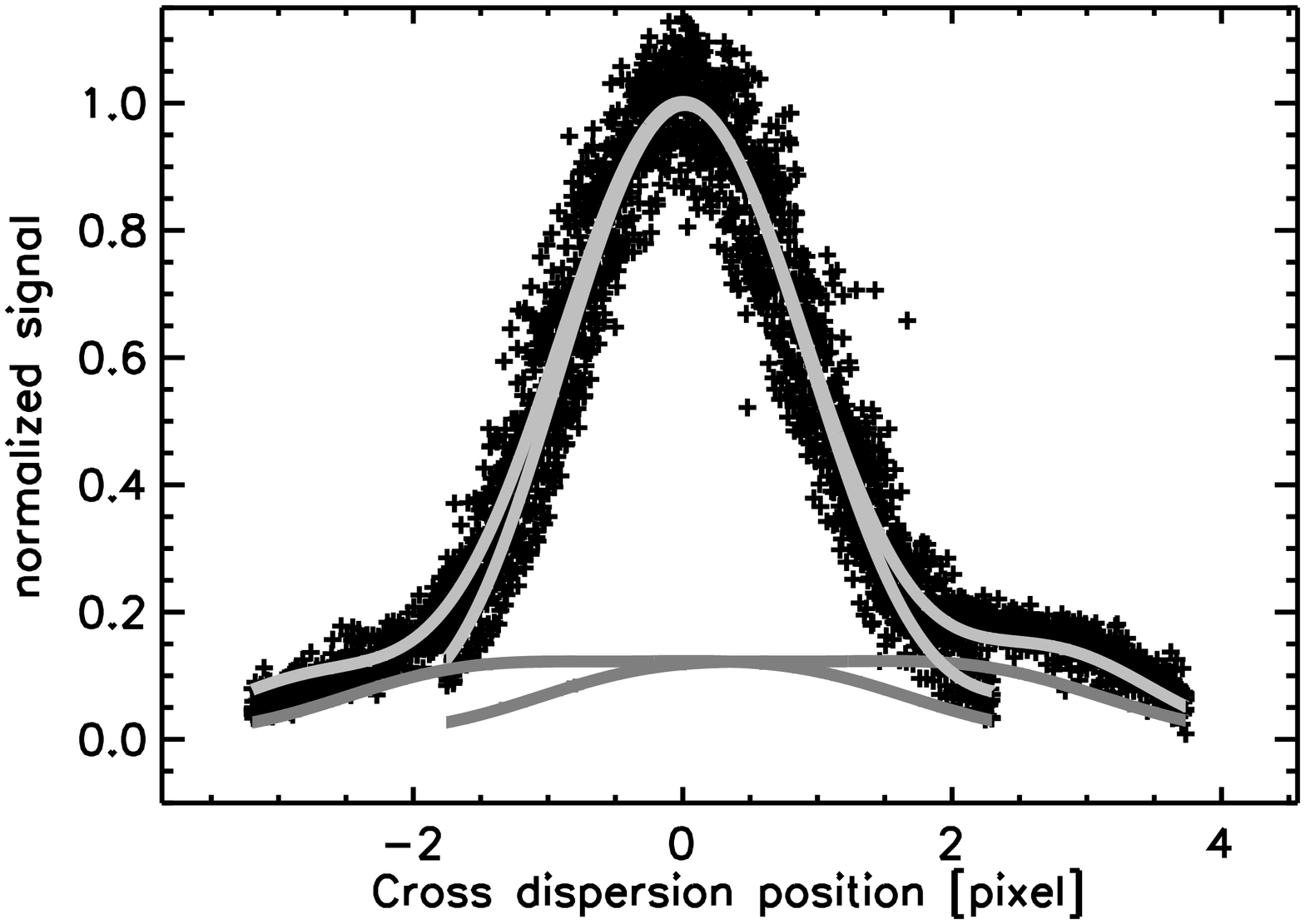}}%
\parbox{0.45\textwidth}{\includegraphics[width=0.45\textwidth]{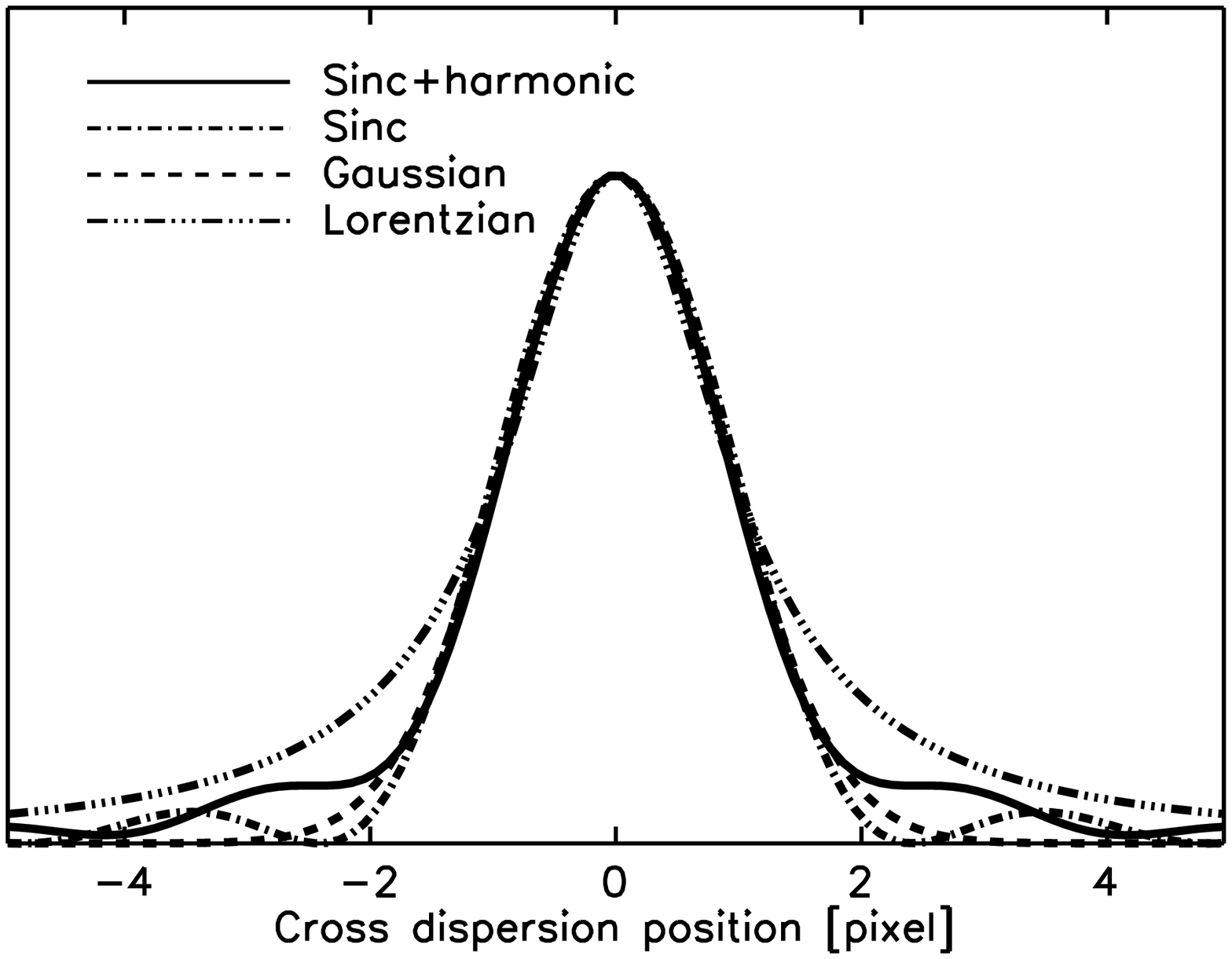}\newline}
\caption{\small\label{fig:xdisp-profile}%
{\small
Illustration of the \textit{Spitzer} IRS cross dispersion profile used in the optimal extraction (see Sec. \ref{sec:optimal-extraction}). The left plot shows a fit to the data of an IRS SH order of a pointsource with a moderate but clear sky component in the IRS spectra. The SH pixel size is $\sim$2.3 arcsec. For LH the profile is similar but the pixel size is $\sim$2 times larger.
%The echelle data (before flatfielding) of both dither positions are combined, normalized, collapsed along the dispersion direction, and corrected for the cross dispersion dither offsets. 
The observed data are plotted with black plusses. Overplotted are the total profile fit (source plus extended emission) and the included extended component. The shape of the extended emission reflects the IRS flatfield of the (for this source) assumed uniform extended emission. The right plot shows a comparison of an IRS PSF profile (Sinc plus harmonics) for a pointsource compared to the profiles of an undistorted Sinc, a Gaussian, and a Lorentzian profile with the same FWHM. Note the significant variation in the strength and shape of the profile wings. The correct characterization of both the width and the wings of the profile for all IRS orders is essential for extracting the proper source and sky spectra. 
}
}
\end{center}
\end{figure*}

%% file: f3.tex
\begin{figure*}[t]
\begin{center}
\includegraphics[width=0.5\textwidth]{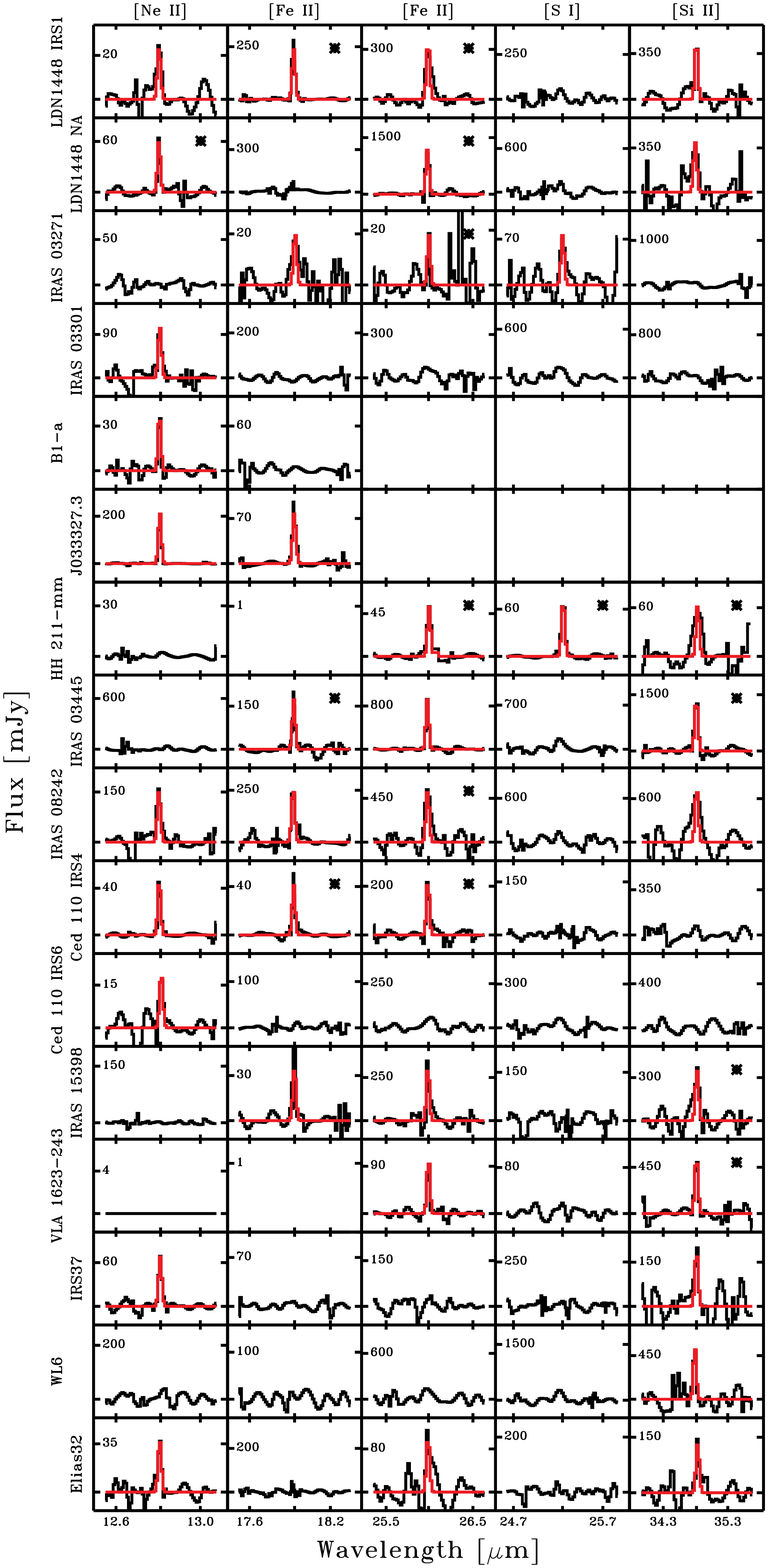}%\hspace{0.04\columnwidth}%
\includegraphics[width=0.5\textwidth]{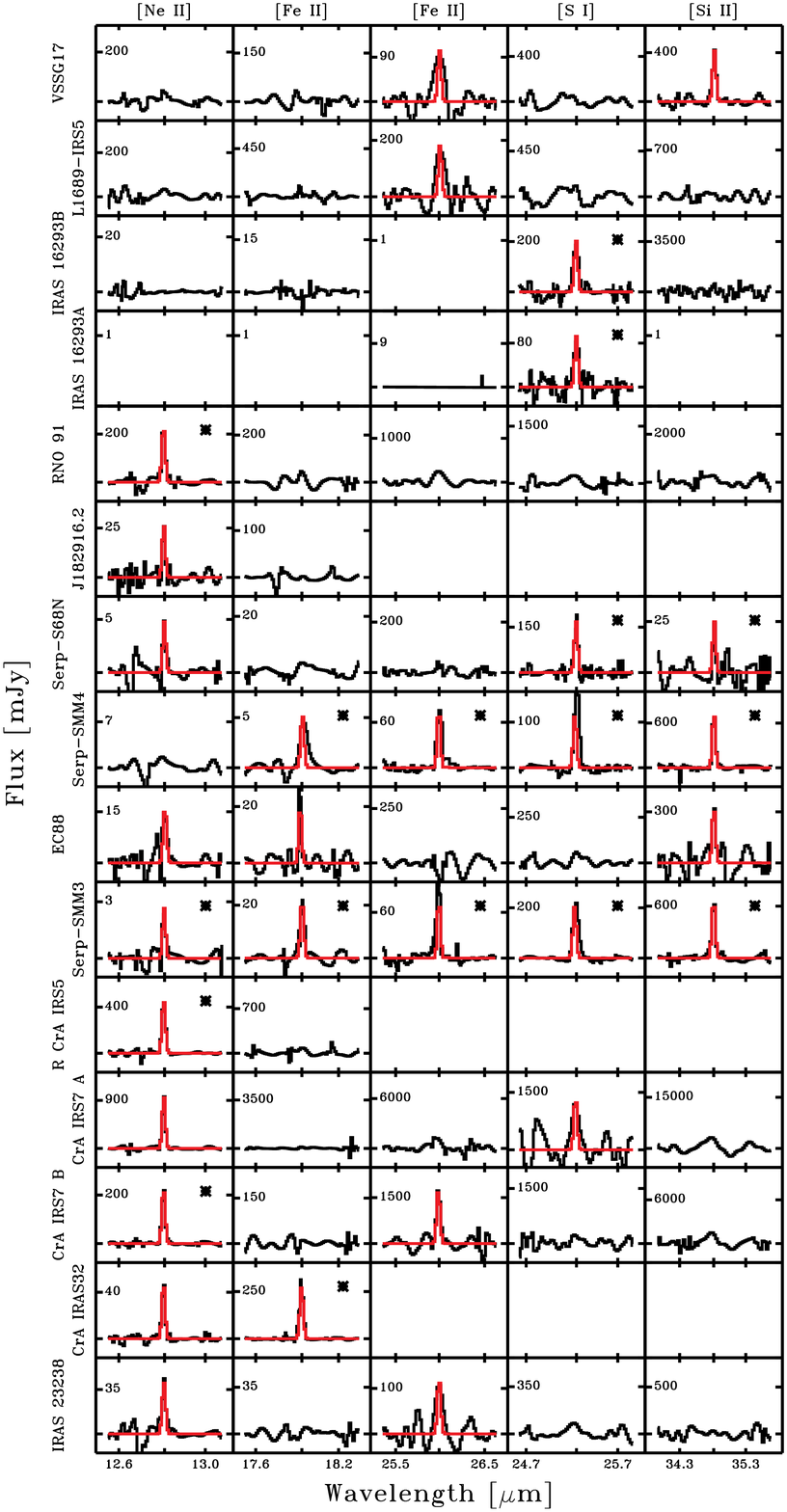}%
\caption{\small\label{fig:atomic_lines}%
Detections of [Ne\,II], [Fe\,II], [S\,I] 
and [Si\,II] at the $\geq4\sigma$ level toward the c2d sample of 
embedded YSOs. Overplotted in red is a Gaussian fit to the observed spectrum.
% and T~Tauri stars with edge-on disks.
Sources indicated with a star show evidence of a (partial) 
contribution of extended emission to the total line flux.
}
\end{center}
%\vspace{14pc}
\end{figure*}

%% file: f4.tex
\begin{figure*}
\vspace{2pc}
\begin{center}
\includegraphics[width=0.5\textwidth]{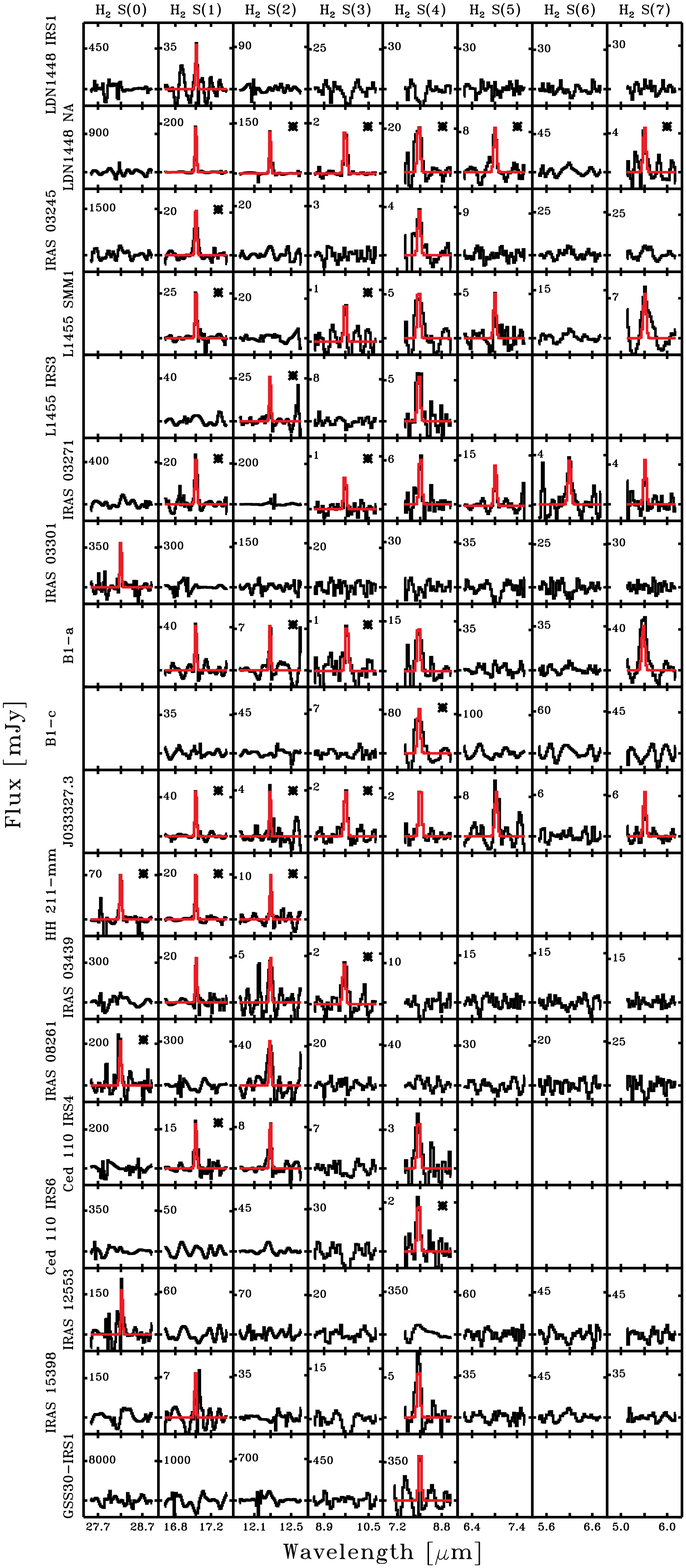}%\hspace{0.04\columnwidth}%
\includegraphics[width=0.5\textwidth]{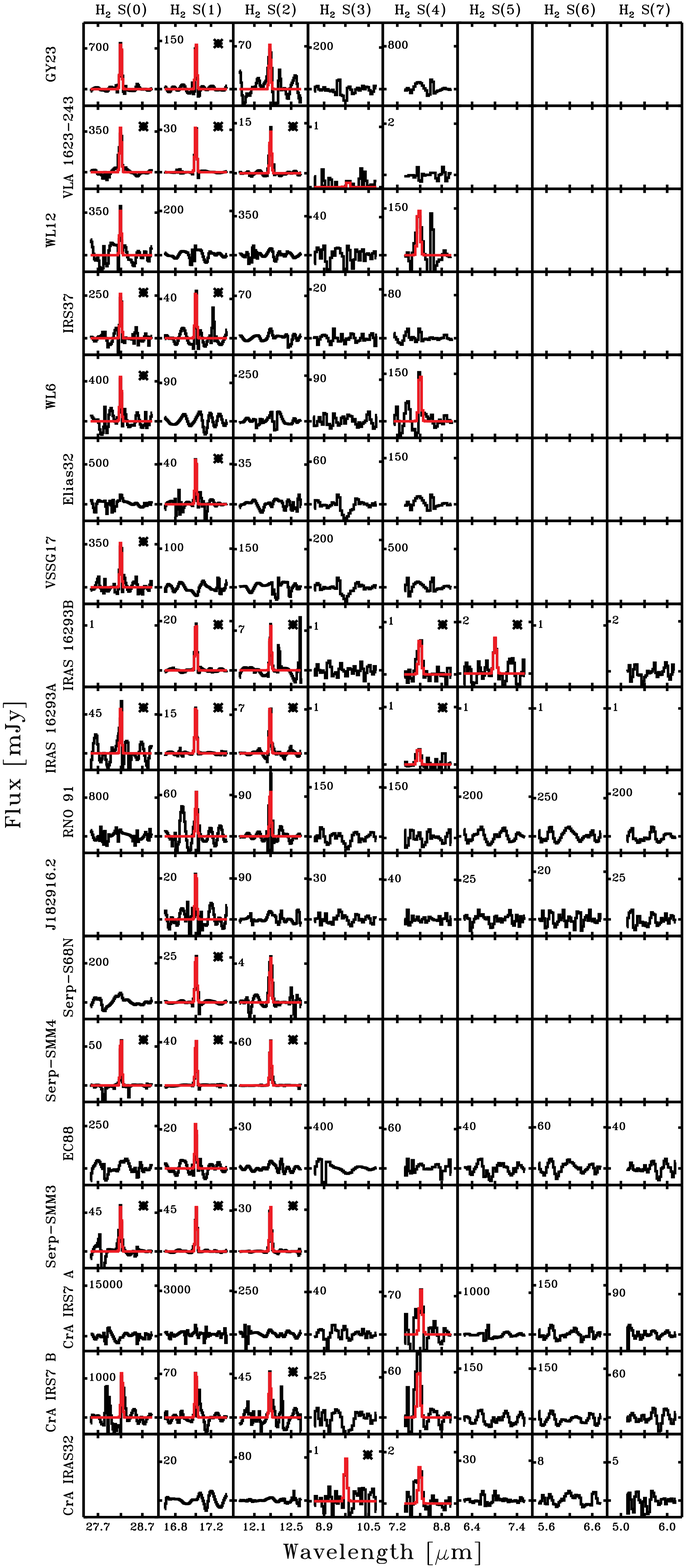}%
\caption{\small\label{fig:H2_lines}%
Detections of H$_2$ emission lines at the $\geq 4\sigma$ level toward the 
c2d sample of embedded YSOs.
Overplotted in red is a Gaussian fit to the observed spectrum.
Sources indicated with a star show evidence of a (partial) 
contribution of extended emission to the total line flux.
}
\end{center}
\vspace{4pc}
\end{figure*}

%% file: f5-6.tex
%\begin{minipage}{\textwidth}
\begin{figure*}[t]
\begin{minipage}[t]{0.5\textwidth}
\begin{center}
\includegraphics[width=0.95\columnwidth]{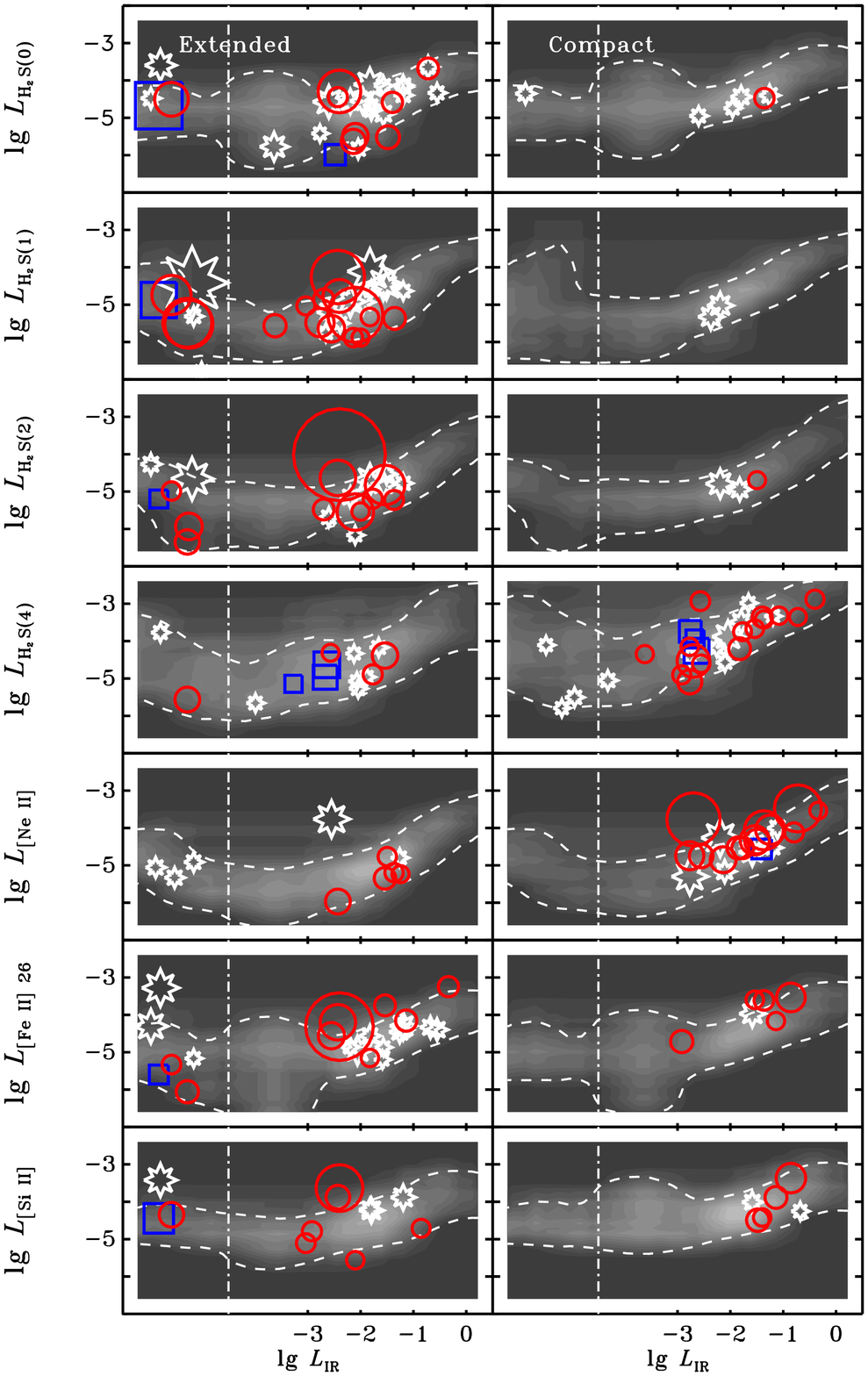}
\begin{minipage}[t]{0.9\columnwidth}
\caption{\small\label{fig:line_relations}%
Line luminosities and upper limits (in solar luminosity) as functions of
the mid-IR luminosity (calculated from the IRS spectra) for the estimated
extended and compact source contributions to the total line strength.
The red circles show the truly embedded sources, white stars the disk sources
\citep[this paper and][]{lahuis07c} and blue squares the unclassified sources.
Shown are all $4\sigma$ line detections, the symbol size scales with line error.
The observed upper limits listed in Tables \ref{atomic:tab:linefluxes} and \ref{h2:tab:linefluxes} are shown as grayscale images with dashed contours. The upper limits are presented as number density distributions with the contours bracketing the 10 and 90\,\% ranges.
All sources plotted left of the vertical dashed line either have no
detected continuum in the IRS spectra or the spectrum does not cover
the mid-IR range (x-position offset for clarity).
}%
\end{minipage}%
\end{center}%
\end{minipage}%
%No clear relation is found for any of these. The apparent 
%relation with the mid-IR luminosity is possoibly due to 
%a detection selection because of noise limits.
%\end{center}
%\end{figure}
%\begin{figure}[H]
%\begin{center}
\begin{minipage}[t]{0.5\textwidth}
\begin{center}%
\includegraphics[width=0.95\columnwidth]{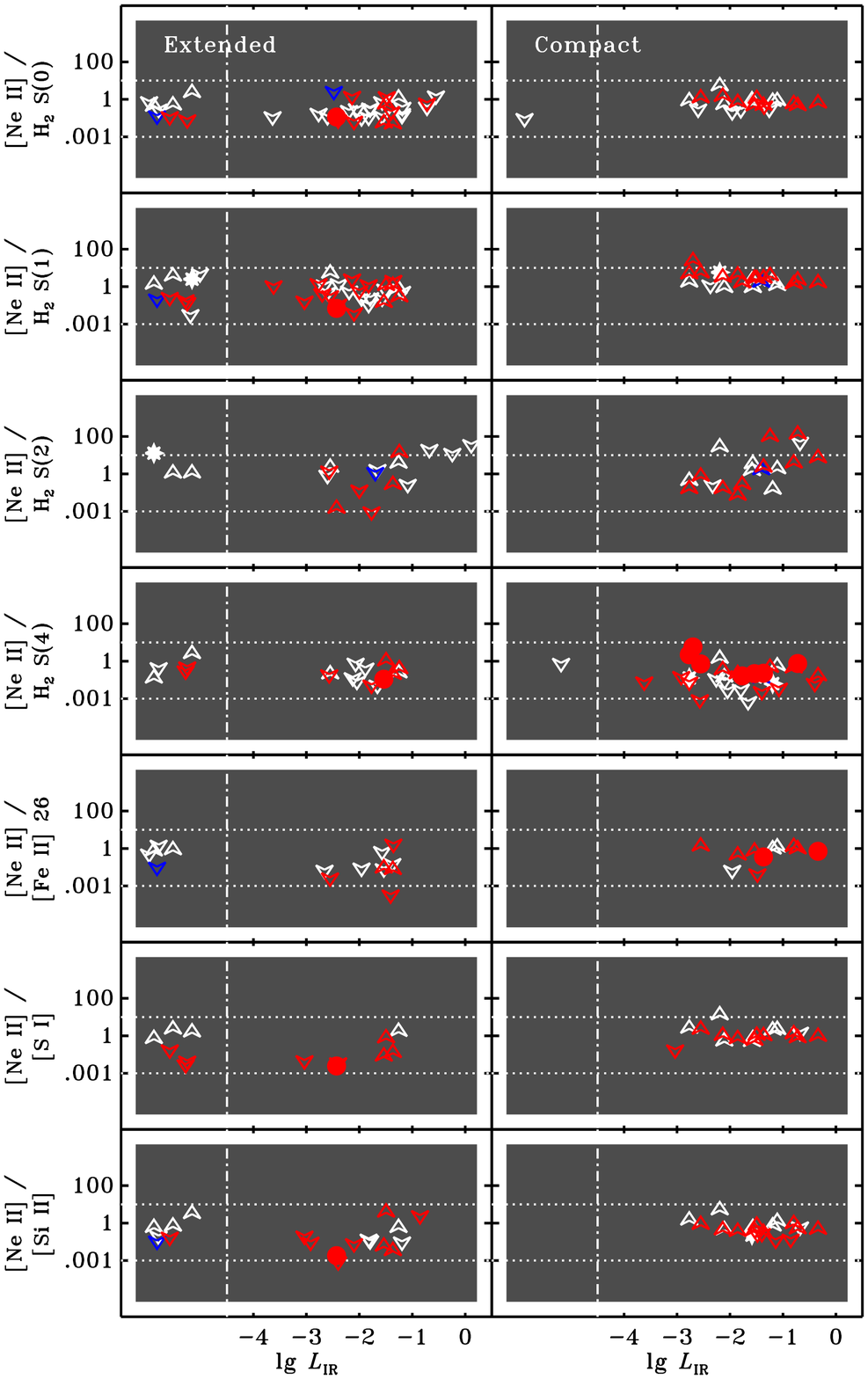}
\begin{minipage}[t]{0.9\columnwidth}
\caption{\small\label{fig:line_ratios}%
Line ratios (filled symbols) and line ratio upper and lower limits of [Ne\,{\sc ii}] with a number of H$_2$ and atomic lines. The use of symbols and colors is the same as in Figure \ref{fig:line_relations}. All sources plotted left of the vertical dashed line either have no detected continuum in the IRS spectra or the spectrum does not cover the mid-IR range (x-position offset for clarity). The two horizontal dotted lines span the range of 0.001 to 10 within which most detections fall. This range is also indicated in Figure \ref{fig:line_ratios_shock} which shows \neii{} line ratios from shock models by \citet{hollenbach89}. Note that most line ratios diplayed here are upper and lower limits. Only for \neii/\htwo\,S(4) a reasonable number of line ratios (filled red circles) exist.
}%
\end{minipage}%
\end{center}
\end{minipage}%
\end{figure*}

%% file: f7.tex
\begin{figure}[]
\begin{center}
\includegraphics[width=\columnwidth]{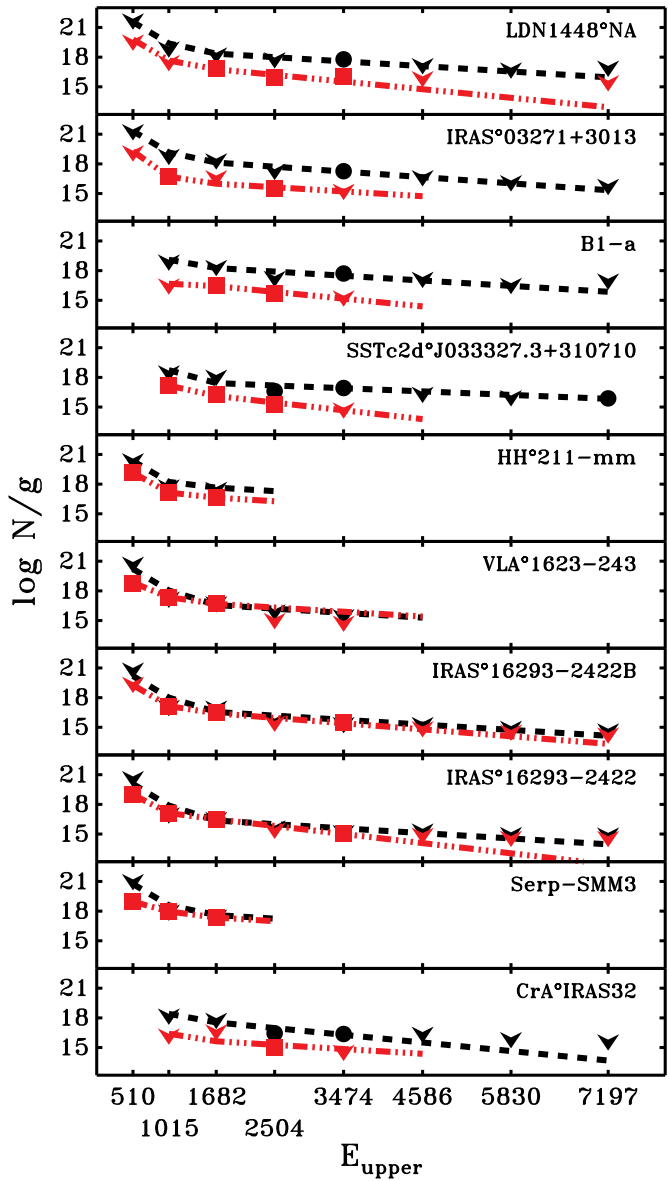}%
\caption{\label{fig:ex_diagram}%
Excitation diagrams of the observed \htwo{} emission. Plotted with 
black circles is the observed spatially compact emission and with
red squares the spatially extended emission. 
The arrows are the derived upper limits. Overplotted with
respectively dashed and dashed/dotted lines is a two temperature
fit to the observed \htwo{} emission,
as seen by the difference in slope going from the low to 
the high energy states. The inferred temperatures are listed in
Table \ref{embtab:gasparam}. Note that the low temperature component
is not always well constrained.
}%\vspace{2\baselineskip}}
\end{center}
\end{figure}

%% file: f8.tex
\begin{figure}[!t]
\begin{center}
\includegraphics[width=0.9\columnwidth]{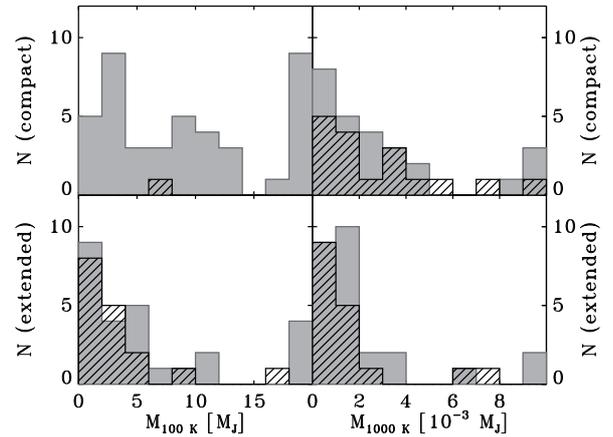}
\caption{\label{fig:h2_mass}\small%
Distribution of H$_2$ masses (in Jupiter mass) for both the compact
and the extended component assuming $T_\mathrm{ex}=100$\, and 1000\,K.
The gray bars represent upper limits on the gas mass, while the hatched
bars indicate sources where H$_2$ is detected at $4\sigma$ or more.
For 100\,K this includes sources with the H$_2$ S(0) and/or S(1) line detected
and for 1000\,K sources with any of the higher $J$ lines detected.
The highest bar also includes all sources with masses higher than
the upper plot limit.
}
\end{center}
\end{figure}

%% file: f9.tex
\begin{figure}[t]
\begin{center}
\includegraphics[width=\columnwidth]{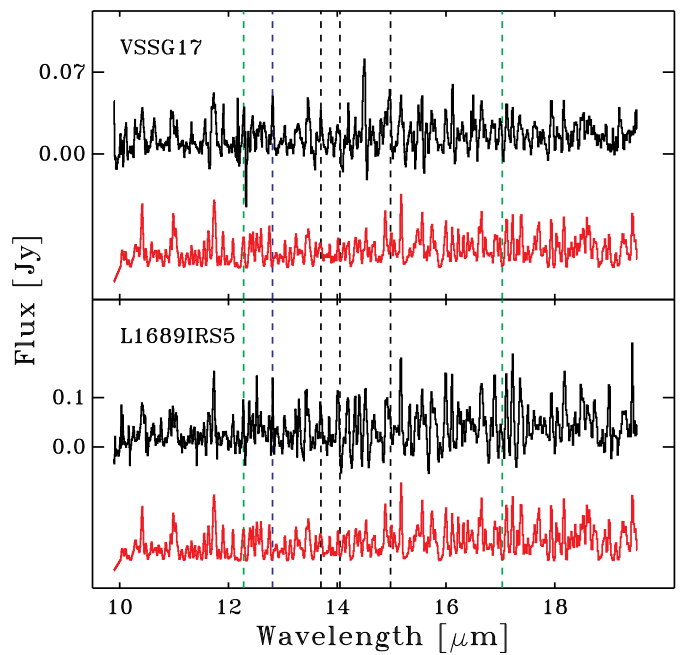}
\caption{\small\label{fig:water}%
Molecular emission toward two truly embedded sources. Plotted for each source are the continuum subtracted \spitzer{} IRS-SH spectra (top spectrum in black) and a single temperature LTE water model at \hbox{$T_{\mathrm ex}=900$\,K} (bottom spectrum in red). Vertical lines indicate the positions of a few important spectral features; from left to right \htwo\,S(2), \neii, \acetylene, HCN, \cotwo{} and \htwo\,S(1). In VSSG17 \cotwo{} and tentative \acetylene{} and HCN emission can be seen.
}
\end{center}
\end{figure}

%% file: f10.tex
\begin{figure*}
\begin{center}
\noindent%
\includegraphics[width=0.475\textwidth]{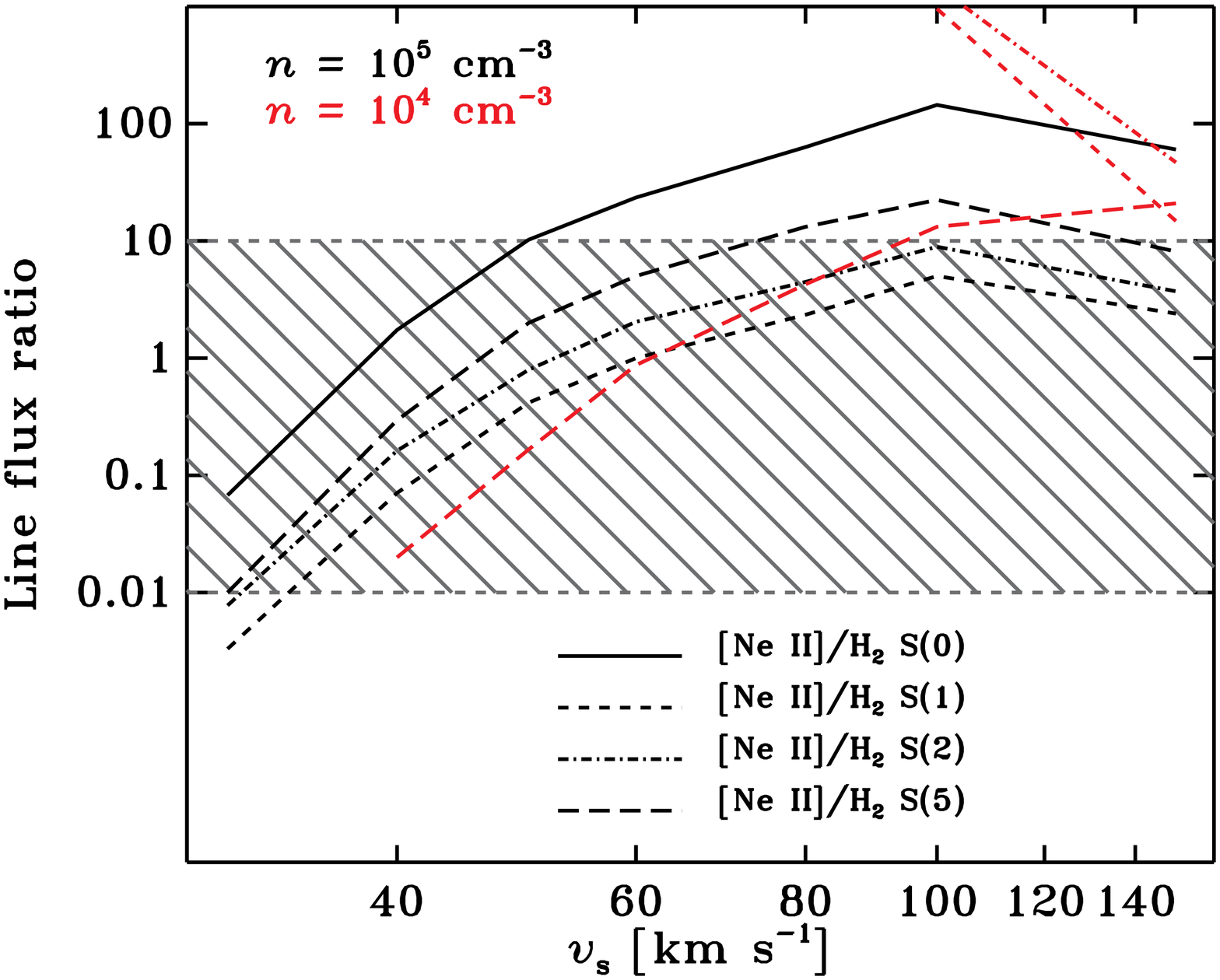}%
\hspace{0.045\textwidth}%
\includegraphics[width=0.475\textwidth]{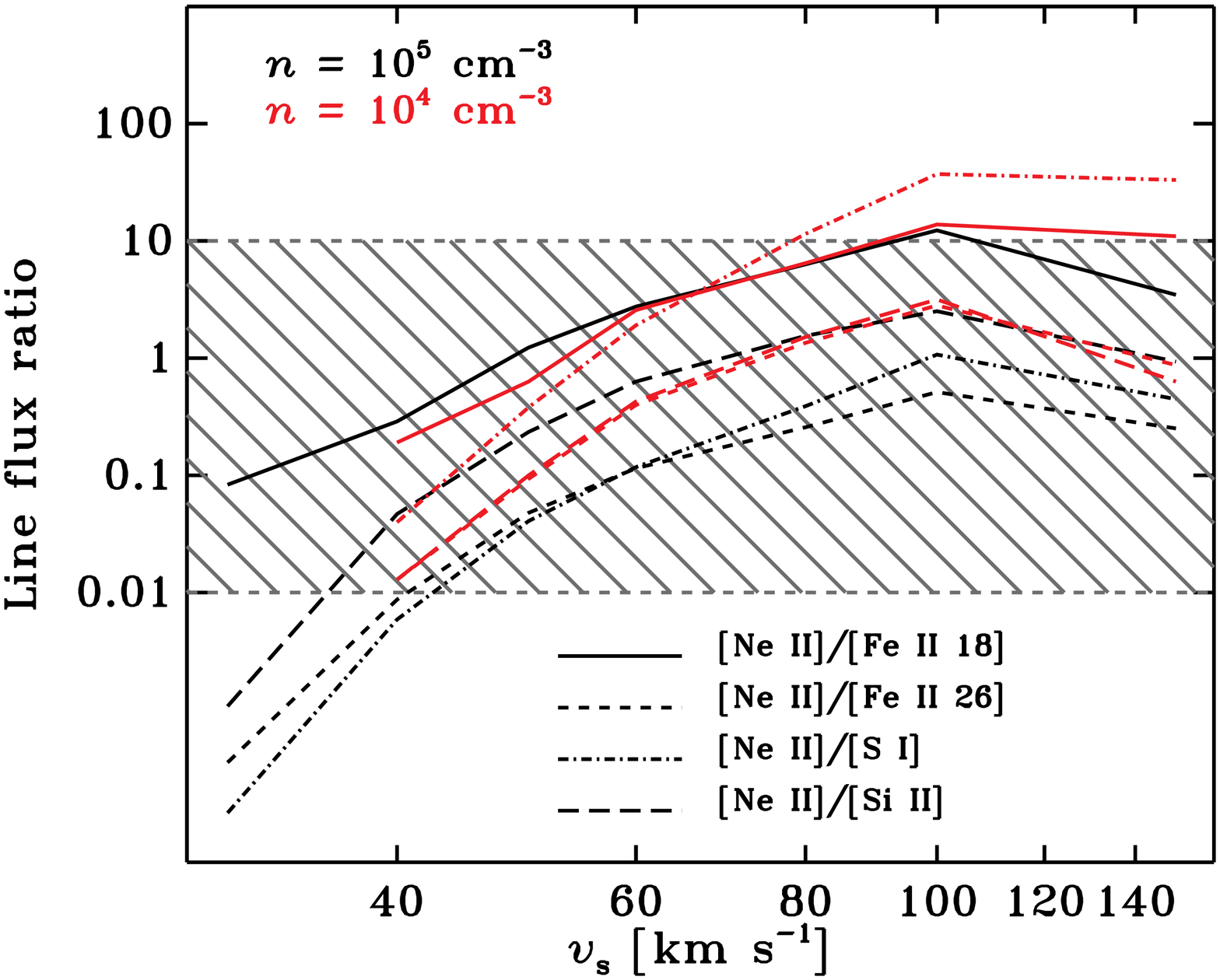}%
\caption{\small\label{fig:line_ratios_shock}%
[Ne\,II] line ratios from $J$-shock models by 
\citet[][]{hollenbach89} for pre-shock densities of $10^5$ (black)
and $10^4$ (red). Note that for the low density case the \neii{} ratios
with the low-$J$ \htwo{} lines are large and quickly fall off the plot.
The hatched area indicates the range of observed 
line ratios (see Figure \ref{fig:line_ratios}).
}%\vspace{1\baselineskip}
\end{center}
\end{figure*}
%\end{figure*}

%% file: f11.tex
\begin{figure}[hb]
\begin{center}
\includegraphics[width=\columnwidth]{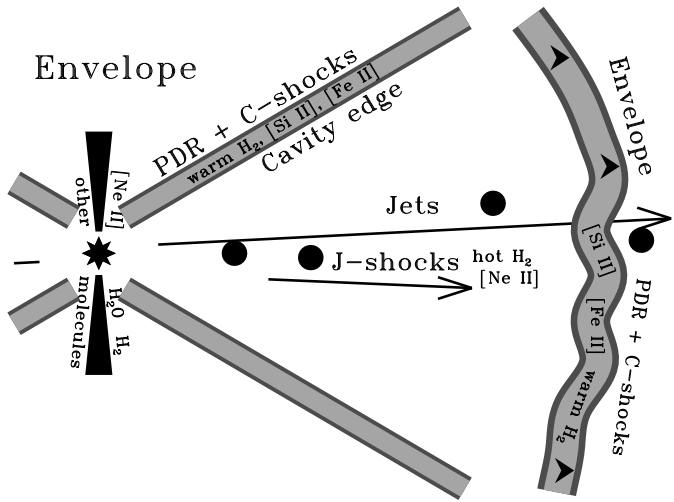}%
\caption{\small\label{fig:cartoon}%
Cartoon image to illustrate the various emission components
for the embedded sources as seen by mid-IR \spitzer-IRS observations.
Extended emission may be observed as PDR and/or \cshock{} emission
from the irradiated and accelerated gas in the cavity and envelope
surfaces. Compact \jshock{} emission can be observed from shocks
within high-velocity jets, indicated by filled circles.
Disk emission may be observed for systems viewed along the outflow 
cavity.
%Figure 11.   In caption mention what the black filled circles
%>        are.  I think they are blobs in the jets that are shocked by the jet.
%>        The jet is likely to blast through the envelope, whereas the wider
%>        wind may cause the kind of shocked shell that you show.  You might
%>        sketch that this shell will be moving outwards and that the wind shocks
%>        it on the inside and then the shell sweeps up the envelope on the
%>        outside with another shock.
}
\end{center}
\end{figure}

%% file: tab1.tex
\begin{table*}[t]
\caption{Source list\label{embtab:sourcelist} }
\scriptsize
\begin{tabular}{rlccccc}
\multicolumn{7}{c}{{\normalsize Embedded sources}} \\
\hline
\hline
{\large\strut} {\#} & Source\hfill
                    & RA
                    & Dec
                    & AOR
                    & tint [s]
                    & D\tablenotemark{a}
                    \\
{\large\strut} & 
               & 
               & 
               & 
               & (SL/SH/LH)*ndith.
               & [pc]
               \\
\hline
\T       1 & LDN1448\ IRS1 & \hfill\RA{3}{25}{09}{.4} & \hfill\Dec{30}{46}{21}{.7}\hfill & 5656832\hfill & (14*1/31*2/60*1)*2 & 250 \\
       2 & LDN1448\ NA & \hfill\RA{3}{25}{36}{.5} & \hfill\Dec{30}{45}{21}{.2}\hfill & 5828096\hfill & (60*1/121*2/60*1)*2 & 250 \\
       3 & IRAS\ 03245+3002 & \hfill\RA{3}{27}{39}{.0} & \hfill\Dec{30}{12}{59}{.4}\hfill & 6368000\hfill & (14*4/31*4/60*1)*2 & 250 \\
       4 & L1455\ SMM1 & \hfill\RA{3}{27}{43}{.2} & \hfill\Dec{30}{12}{28}{.8}\hfill & 15917056\hfill & (60*2/121*2/--)*2 & 250 \\
       5 & L1455\ IRS3 & \hfill\RA{3}{28}{00}{.4} & \hfill\Dec{30}{08}{01}{.3}\hfill & 15917568\hfill & (14*4/121*2/--)*2 & 250 \\
\end{tabular}
\tablenotetext{a}{Assumed cloud distances;
                  Chamaeleon \citep[178\,pc][]{whittet97},
                  Lupus I, IV (150\,pc) and Lupus III \citep[200\,pc][]{comeron08},\\
                  Ophiuchus \citep[125\,pc][]{degeus89},
                  Perseus \citep[250\,pc, see discussion in][]{enoch06},
                  Taurus-Auriga \citep[160\,pc][]{kenyon94},\\
                  Serpens \citep[260\,pc][]{straizys96}
                 }
\end{table*}

%% file: tab2.tex
\begin{table*}[t]
\caption{Source charateristics\label{embtab:features}}
\scriptsize
\begin{tabular}{rlccccccccc}
\multicolumn{10}{c}{{\normalsize Embedded sources}} \\
\hline
\hline
{\large\strut}
           {\#}
         & {Source}\hfill
         & Warm H$_2$
         & Hot H$_2$
         & \neii
         & [Fe\,{\sc ii}]
         & [Si\,{\sc ii}]
         & H$_2$O\tablenotemark{a}
         & outflow\tablenotemark{b}
         & source size
         \\ 
{\large\strut}
           {}
         & {}\hfill
         & ext.
         & src
         & src
         & ext.
         & ext.
         & 
         & 
         & [\arcsec]
         \\ 
\hline
 1 & LDN1448\ IRS1         &          &          &          &  \cmrk   &          &          & y  & -- \\
 2 & LDN1448\ NA           &  \cmrk   &  \cmrk   &  \cmrk   &  \cmrk   &          &          & y  & -- \\
 3 & IRAS\ 03245+3002      &  \cmrk   &  \cmrk   &          &          &          &          & y  &    3.1(1.5) \\
 4 & L1455\ SMM1           &  \cmrk   &  \cmrk   &          &          &    --    &          & n  &    0.2(0.2) \\
 5 & L1455\ IRS3           &  \cmrk   &          &          &          &    --    &  \cmrk   & n  & -- \\
\end{tabular}
\tablenotetext{a}{One \cmrk indicates a detection in one extraction; two \cmrk\cmrk indicates sources with detection in both the optimal extraction and the full aperture extraction.}
\tablenotetext{b}{Evidence for the presence of outflows based on
a visual inspection of \spitzer-IRAC images;
\textbf{y}es, \textbf{n}o or \textbf{c}onfused.}
\end{table*}

%% file: tab3_1.tex
\begin{table*}[t]
\caption{Observed linefluxes and $1\sigma$ uncertainties ($\mathrm{10^{-16} erg\,cm^{-2}\,s^{-1}}$)\label{atomic:tab:linefluxes}}
\scriptsize
\begin{tabular}{rlrrrrrrrrrrrr}
\multicolumn{14}{c}{{\normalsize Embedded sources}} \\
\hline
\hline
{\large\strut}
{\#} & {Source}\hfill 
 & & \multicolumn{3}{c}{[Ne II]} 
 & & \multicolumn{3}{c}{[Ne III]} 
 & & \multicolumn{3}{c}{[Fe I] 24.04} 
         \\
\cline{4-6}
\cline{8-10}
\cline{12-14}
  {\large\strut}  & {} 
 & {} & src\tablenotemark{a} & ext\tablenotemark{a} & upp\tablenotemark{a}
 & {} & src & ext & upp
 & {} & src & ext & upp
 \\
\hline
\T 1 & LDN1448\ IRS1        & {} &   --  &   --  & ($   30/   23$)  & {} &   --  &   --  & ($   43/    7$)  & {} &   --  &   --  & ($  110/   44$)  \\
 2 & LDN1448\ NA          & {} &   240 &    23 & ($   20/    3$)  & {} &   --  &   --  & ($   40/   28$)  & {} &   --  &   --  & ($  140/   99$)  \\
 3 & IRAS\ 03245+3002     & {} &   --  &   --  & ($    6/    5$)  & {} &   --  &   --  & ($   20/   20$)  & {} &   --  &   --  & ($  110/   60$)  \\
 4 & L1455\ SMM1          & {} &   --  &   --  & ($   10/    8$)  & {} &   --  &   --  & ($    9/    2$)  &       &       &           &  \\
 5 & L1455\ IRS3          & {} &   --  &   --  & ($   18/    2$)  & {} &   --  &   --  & ($   15/    2$)  &       &       &           &  \\
\end{tabular}
\tablenotetext{a}{\textbf{src} = compact (or slightly extended, 
                  see Table \ref{embtab:features}) component
                  centered on source ; 
                  \textbf{ext} = fully extended (to \spitzer-IRS)
                  component ; 
                  \textbf{upp} = $1\sigma$ upper limit of src/ext linefluxes}
\end{table*}

%% file: tab3_2.tex
\begin{table*}[t]
\addtocounter{table}{-1}
\caption{ -- continued --}
\scriptsize
\begin{tabular}{rrrrrrrrrrrrrrrrr}
\multicolumn{17}{c}{{\normalsize Embedded sources}} \\
\hline
\hline
{\large\strut}
{\#} 
 & & \multicolumn{3}{c}{[Fe II] 17.94} 
 & & \multicolumn{3}{c}{[Fe II] 25.99} 
 & & \multicolumn{3}{c}{[S I]} 
 & & \multicolumn{3}{c}{[Si II]} 
         \\
\cline{3-5}
\cline{7-9}
\cline{11-13}
\cline{15-17}
  {\large\strut} 
 & {} & src & ext & upp
 & {} & src & ext & upp
 & {} & src & ext & upp
 & {} & src & ext & upp
 \\
\hline
 1 & {} &   370 &   --  & ($   91/   20$)  & {} &   350 &   360 & ($   83/   57$)  & {} &   --  &   --  & ($   63/   43$)  & {} &   670 &   --  & ($   99/   57$)  \\
 2 & {} &   --  &   --  & ($   50/   12$)  & {} &  1300 &   920 & ($  300/  150$)  & {} &   --  &   --  & ($  150/  100$)  & {} &   --  &   --  & ($  210/  140$)  \\
 3 & {} &   --  &   --  & ($   28/   25$)  & {} &   --  &   --  & ($  110/   75$)  & {} &   --  &   --  & ($  140/   83$)  & {} &   --  &   --  & ($  390/   30$)  \\
 4 & {} &   --  &   --  & ($   24/    5$)  &       &       &           &  &       &       &           &  &       &       &           &  \\
 5 & {} &   --  &   --  & ($   19/    4$)  &       &       &           &  &       &       &           &  &       &       &           &  \\
\end{tabular}
\end{table*}

%% file: tab4_1.tex
\begin{table*}[t]
\caption{Observed linefluxes and $1\sigma$ uncertainties ($\mathrm{10^{-16} erg\,cm^{-2}\,s^{-1}}$)\label{h2:tab:linefluxes}}
\scriptsize
\begin{tabular}{rlrrrrrrrrrrrrrrrr}
\multicolumn{18}{c}{{\normalsize Embedded sources}} \\
\hline
\hline
{\large\strut}
{\#} & {Source}\hfill 
 & & \multicolumn{3}{c}{H2 S(0)} 
 & & \multicolumn{3}{c}{H2 S(1)} 
 & & \multicolumn{3}{c}{H2 S(2)} 
 & & \multicolumn{3}{c}{H2 S(3)} 
         \\
\cline{4-6}
\cline{8-10}
\cline{12-14}
\cline{16-18}
  {\large\strut}  & {} 
 & {} & src\tablenotemark{a} & ext\tablenotemark{a} & upp\tablenotemark{a}
 & {} & src & ext & upp
 & {} & src & ext & upp
 & {} & src & ext & upp
 \\
\hline
\T 1 & LDN1448\ IRS1        & {} &   --  &   --  & ($  110/   71$)  & {} &   --  &   --  & ($   51/   45$)  & {} &   --  &   --  & ($   42/   37$)  & {} &   --  &   --  & ($  140/   35$)  \\
 2 & LDN1448\ NA          & {} &   --  &   --  & ($  190/  150$)  & {} &   --  &   --  & ($   31/   58$)  & {} &   --  &    77 & ($   26/    3$)  & {} &   --  &    76 & ($  130/    4$)  \\
 3 & IRAS\ 03245+3002     & {} &   --  &   --  & ($  240/  130$)  & {} &   --  &    14 & ($   11/    2$)  & {} &   --  &   --  & ($   14/   12$)  & {} &   --  &   --  & ($   13/    4$)  \\
 4 & L1455\ SMM1          &       &       &           &  & {} &   --  &    17 & ($    8/    1$)  & {} &   --  &   --  & ($   14/   12$)  & {} &   --  &    14 & ($    7/    3$)  \\
 5 & L1455\ IRS3          &       &       &           &  & {} &   --  &   --  & ($   18/    2$)  & {} &   --  &    15 & ($   25/    3$)  & {} &   --  &   --  & ($   33/   22$)  \\
\end{tabular}
\tablenotetext{a}{see footnote $a$ of Table \ref{atomic:tab:linefluxes}}
\end{table*}

%% file: tab4_2.tex
\begin{table*}[t]
\addtocounter{table}{-1}
\caption{ -- continued --}
\scriptsize
\begin{tabular}{rrrrrrrrrrrrrrrrr}
\multicolumn{17}{c}{{\normalsize Embedded sources}} \\
\hline
\hline
{\large\strut}
{\#} 
 & & \multicolumn{3}{c}{H2 S(4)} 
 & & \multicolumn{3}{c}{H2 S(5)} 
 & & \multicolumn{3}{c}{H2 S(6)} 
 & & \multicolumn{3}{c}{H2 S(7)} 
         \\
\cline{3-5}
\cline{7-9}
\cline{11-13}
\cline{15-17}
  {\large\strut} 
 & {} & src & ext & upp
 & {} & src & ext & upp
 & {} & src & ext & upp
 & {} & src & ext & upp
 \\
\hline
 1 & {} &   --  &   --  & ($  330/   40$)  & {} &   --  &   --  & ($  160/    0$)  & {} &   --  &   --  & ($  160/    0$)  & {} &   --  &   --  & ($  190/    0$)  \\
 2 & {} &  1100 &   210 & ($  250/   23$)  & {} &   --  &   440 & ($  270/   57$)  & {} &   --  &   --  & ($  220/    0$)  & {} &   --  &   350 & ($  670/   80$)  \\
 3 & {} &   230 &   --  & ($   55/   11$)  & {} &   --  &   --  & ($   51/    0$)  & {} &   --  &   --  & ($  100/    0$)  & {} &   --  &   --  & ($  160/    0$)  \\
 4 & {} &   360 &   --  & ($   84/    2$)  & {} &   210 &   --  & ($   34/    0$)  & {} &   --  &   --  & ($   53/    0$)  & {} &   --  &   --  & ($  150/    0$)  \\
 5 & {} &   --  &   --  & ($  100/   12$)  &       &       &           &  &       &       &           &  &       &       &           &  \\
\end{tabular}
\end{table*}

%% file: tab5.tex
\begin{table*}[t]
\caption{Diagnostic parameters of the gas contents\label{embtab:gasparam}}
\scriptsize
\begin{tabular}{rlccccccccccccccc}
\multicolumn{17}{c}{{\normalsize Embedded sources}} \\
\hline
\hline
{\large\strut} {\#}
           & Source
           & \multicolumn{7}{c}{Warm H$_2$\tablenotemark{a}}
           & {}
           & \multicolumn{7}{c}{Hot H$_2$\tablenotemark{a}}
         \\
 \cline{3-9}\cline{11-17}
           & {\large\strut}
           & \multicolumn{3}{c}{Compact\tablenotemark{b}}
           & {}
           & \multicolumn{3}{c}{Extended}
           & {}
           & \multicolumn{3}{c}{Compact\tablenotemark{b}}
           & {}
           & \multicolumn{3}{c}{Extended}
         \\
 \cline{3-5}\cline{7-9}\cline{11-13}\cline{15-17}
             {\large\strut}
           & {}
           & {[K]}
           & {[$10^{22}\mathrm{cm^{-2}}$]}
           & {[$\mathrm{M_J}$]}
           & {}
           & {[K]}
           & {[$10^{20}\mathrm{cm^{-2}}$]}
           & {[$\mathrm{M_J}$]}
           & {}
           & {[K]}
           & {[$10^{20}\mathrm{cm^{-2}}$]}
           & {[$10^{-3}\,\mathrm{M_J}$]}
           & {}
           & {[K]}
           & {[$10^{18}\mathrm{cm^{-2}}$]}
           & {[$10^{-3}\,\mathrm{M_J}$]}
 \\
\hline
\T1 & LDN1448\,IRS1        & $$ & $< 600$ & $< 18.4$ &  & $$ & $<  14$ & $<  7.9$ &  & $$ & $<  10$ & $<  3.2$ &  & $$ & $<   2$ & $<  1.5$ \\
2 & LDN1448\,NA          & $$ & $< 600$ & $< 19.2$ &  & $$ & $<  29$ & $< 17.1$ &  & $$ & $   6$ & $  1.8$ &  & $ 623$ & $   6$ & $  3.8$ \\
3 & IRAS\,03245+3002     & $$ & $< 400$ & $< 12.6$ &  & $$ & $   3$ & $  1.8$ &  & $$ & $   2$ & $  0.9$ &  & $$ & $0.5$ & $  0.3$ \\
4 & L1455\,SMM1          & $$ & $<  80$ & $<  2.5$ &  & $$ & $   1$ & $  1.1$ &  & $$ & $   1$ & $  0.4$ &  & $$ & $0.5$ & $  0.3$ \\
\end{tabular}
\tablenotetext{a}{100\,K is assumed for the warm component
                  and 1000\,K for the hot component.
                  The derived column density and mass of the
                  warm component depend strongly on the assumed
                  temperature. A temperature of 150 and 200\,K
                  reduces the column density and mass by
                  respectively a factor of $\sim$ 30 and 140.
                  For the hot component 1500\,K instead of
                  1000\,K may results in a reduction of column
                  density and mass up to a factor of 10.}
\tablenotetext{b}{For the unresolved emission a source with
                  $r=50$\,AU is assumed to obtain an estimate
                  of the column density.}
\end{table*}